\documentclass{emulateapj}
\usepackage{rotating}
\usepackage{epsf}
\usepackage{psfig}
\usepackage{float}
\usepackage [english]{babel}
\usepackage [autostyle, english = american]{csquotes}
\usepackage{chngcntr}
\MakeOuterQuote{"}
\pdfoutput=1 

\shorttitle{Taffy PACS}
\shortauthors{Peterson et al.}

\newcommand{\Hi}{\ion{H}{1} }

\newcommand{\Ht}{${\rm H_{2}}$ }
\newcommand{\cii}{[\ion{C}{2}] }
\newcommand{\oi}{[\ion{O}{1}] }

\begin{document}

\title{Herschel Spectroscopy of the Taffy Galaxies (UGC 12914/12915 = VV 254): Enhanced [\ion{C}{2}] emission in the collisionally-formed bridge.}

\author{B. W. Peterson\altaffilmark{1}, P. N. Appleton\altaffilmark{2}, T. Bitsakis\altaffilmark{3}, P. Guillard\altaffilmark{4}, K. Alatalo\altaffilmark{5,6},  F. Boulanger\altaffilmark{7}, M. Cluver\altaffilmark{8}, P.-A. Duc\altaffilmark{9}, E. Falgarone\altaffilmark{7},  S. Gallagher\altaffilmark{10}, Y. Gao\altaffilmark{11}, G. Helou\altaffilmark{2}, T. H. Jarrett\altaffilmark{12}, B. Joshi\altaffilmark{13}, U. Lisenfeld\altaffilmark{14}, N. Lu\altaffilmark{15,16}, P. Ogle\altaffilmark{6},      G. Pineau des For\^ets\altaffilmark{17},  P. van der Werf\altaffilmark{18}, C. K. Xu\altaffilmark{15,16}} 

\altaffiltext{1}{University of Wisconsin -- Barron County, 1800 College Dr., Rice Lake, WI 54868, USA. bradley.peterson@uwc.edu}
\altaffiltext{2}{Caltech/IPAC, MC 100-22, 1200 E. California Blvd., Pasadena, CA 91125, USA. apple@ipac.caltech.edu}
\altaffiltext{3}{Instituto de Radioastronom\'ia y Astrof\'isica, Universidad Nacional Aut\'onoma de M\'exico, Morelia, 58089, Mexico} 
\altaffiltext{4}{Sorbonne Universit\'es, UPMC Paris 6 et CNRS, UMR 7095, Institut d'Astrophysique de Paris, 98 bis Bd Arago, 75014 Paris, France}
\altaffiltext{5}{Carnegie Observatories, Carnegie Institute of Washington, Pasadena, CA}
\altaffiltext{6}{Space Telescope Science Institute, 3700, San Martin Drive, Baltimore, MD 21218,  USA}
\altaffiltext{7}{LERMA/LRA, Observatoire de Paris, PSL Research University, CNRS, Sorbonne Universit\'e, 
UPMC Universit\'e Paris 06, Ecole normale sup\'erieure, 75005 Paris, France}
\altaffiltext{8}{Department of Physics and Astronomy, University of the Western Cape, Robert Sobukwe Road, Bellville, 7535, South Africa}
\altaffiltext{9}{Universit\'e de Strasbourg, CNRS, Observatoire astronomique de Strasbourg, UMR 7550, F-67000 Strasbourg, France }
\altaffiltext{10}{Department of Physics and Astronomy, University of Western Ontario, London, ON, N6A 3K7, Canada}
\altaffiltext{11}{Purple Mountain Observatory, Key Lab of Radio Astronomy, Chinese Academy of Sciences, 210008, Nanjing, China}
\altaffiltext{12}{Astronomy Department, University of Cape Town, Rondebosch, 7700, South Africa}
\altaffiltext{13}{School of Earth and Space Exploration, Arizona State University, Tempe, AZ 85287, USA}
\altaffiltext{14}{Departamento de F\'isica Teorica y del Cosmos, Universidad de Granada, Spain and Instituto Carlos I de F\'isica Teorica y Computacional, Facultad de Ciencias, 18071 Granada, Spain}
\altaffiltext{15}{National Astronomical Observatories, Chinese Academy of Sciences (CAS), Beijing 100012, China}
\altaffiltext{16}{China-Chile Joint Center for Astronomy, CAS,  Camino El Observatorio 1515,  Las Condes, Santiago, Chile}
\altaffiltext{17}{IAS (UMR 8617 du CNRS), Universit\'{e} de Paris Sud, F-91405 Orsay, France; LERMA (UMR 8112 du CNRS), Observatoire de Paris, 61 Avenue de l'Observatoire, F-75014 Paris, France}
\altaffiltext{18}{Leiden Observatory, Leiden University, PO Box 9513, 2300 RA Leiden, The Netherlands}


%

\begin{abstract}
Using the PACS and SPIRE spectrometers on-board {\it Herschel}, we obtained observations of the Taffy galaxies (UGC 12914/12915) and bridge. The Taffy system is believed to be the result of a face-on collision between two gas-rich galaxies, in which the stellar disks passed through each other, but the gas was dispersed into a massive \ion{H}{1} and molecular bridge between them. Emission is detected and mapped in both galaxies and the bridge in the [\ion{C}{2}]157.7~$\micron$ and [\ion{O}{1}]63.2~$\micron$ fine-structure lines. Additionally, SPIRE FTS spectroscopy detects the [\ion{C}{1}] $^3$P$_2$$\rightarrow$$^3$P$_1$(809.3~GHz) and [\ion{C}{1}] $^3$P$_1$$\rightarrow$$3$P$_0$(492.2~GHz) neutral carbon lines, and weakly detects high-J CO transitions in the bridge. These results indicate that the bridge is composed of a warm multi-phase medium consistent with shock and turbulent heating. Despite low star formation rates in the bridge, the [\ion{C}{2}] emission appears to be enhanced, reaching [\ion{C}{2}]/FIR ratios of 3.3$\%$ in parts of the bridge. Both the [\ion{C}{2}] and [\ion{O}{1}] lines show broad intrinsic multi-component profiles, similar to those seen in previous CO 1-0 and \ion{H}{1} observations. The [\ion{C}{2}] emission shares similar line profiles with both the double-peaked \ion{H}{1} profiles and shares a high-velocity component with single-peaked CO profiles in the bridge, suggesting that the [\ion{C}{2}] emission originates in both the neutral and molecular phases. We show that it is feasible that a combination of turbulently heated H$_2$ and high column-density \ion{H}{1}, resulting from the galaxy collision, is responsible for the enhanced [\ion{C}{2}] emission.  
\end{abstract}


\section{Introduction} 
The Taffy galaxies (UGC 12914/5) are believed to be the result of a recent (25--30~Myr) face-on collision between two gas-rich disk galaxies, during which the \Hi and molecular clouds in the two disks strongly interacted, leading to the formation of a radio continuum and gas bridge between the visible galaxies \citep{condon1993, struck1997, vollmer2012}. The bridge region contains a significant fraction of the system's \Hi and molecular gas \citep{braine2003, gao2003}, and is also detected through its dust emission \citep{jarrett1999, zhu2007}. It is likely the system contains gas with high levels of turbulence, and so the galaxies and bridge represent an interesting opportunity to study the heating and cooling processes in highly disturbed intergalactic gas. Similar processes may be present at high-$z$, when gas is rapidly accreted onto dark-matter halos. It is therefore important to understand nearby examples of highly turbulent systems.  

Several lines of evidence show that the gas in the Taffy bridge is highly disturbed. CO~(1--0) observations showed broad line widths (FWHM $\sim$200~km~s$^{-1}$) between the galaxies \citep{gao2003}. \citet{zhu2007} found that the strength of the CO~(3--2) and (2--1) lines in the bridge was consistent with warm turbulent gas. Recent ALMA CO observations of bridge gas on sub-arcsec scales show a tangled web of narrow filaments with peculiar kinematics, suggestive of turbulent motions in the bridge (Appleton et al., in preparation).

Evidence for shock heating of the molecular gas also comes from the direct detection of large quantities ($>$~4.2 $\times 10^{8}$~M$_{\odot}$) of warm molecular hydrogen in the bridge through {\it Spitzer} observations of the pure rotational \Ht emission lines \citep{peterson2012}. This study also showed high values of the ratios $L$(H$_2$)/$L$(PAH) and $L$(H$_2$)/$L$(FIR), inconsistent with photoelectric heating from photodissociation regions (PDRs) near star formation sites.  

There is further evidence for shocks and turbulence in much lower density gas. Soft X-ray emission is detected in the northern part of the bridge with {\it Chandra} \citep{appleton2015}. The emission is consistent with million-degree hot diffuse gas heated by fast shocks. Optical emission lines from ionized gas across the bridge have also been detected in a recent optical IFU study of the Taffy (Joshi et al., in preparation). These observations show evidence of multiple-line profiles, and LINER-like excitation consistent with gas heated by 200--300~km~s$^{-1}$ shocks. 

In many ways, the Taffy bridge region shows striking similarities to another spatially-resolved, multi-phased region: the  intergalactic filament in the Stephan's Quintet compact group (hereafter SQ; \citealt{sulentic01, xu03, trinchieri05, osullivan09}). Like the Taffy, the SQ filament contains significant quantities of warm molecular gas that can be modeled by shocks and turbulent energy dissipation resulting from a galaxy collision \citep{appleton2006, cluver2010}. Though \Ht molecules are expected to be destroyed in high-speed shocks, the models of \citet{guillard2009} demonstrated that the molecules can reform on short time scales on dust grains that survive in high-density clumps within a multi-phase shocked medium \citep[mechanical heating is also suspected in translucent clouds in the Galaxy;] []{Ingalls2011}. In a recent analysis, \citet{appleton2017} showed that the excitation of the rotational H$_2$ lines in SQ could be modeled by a combination of low-velocity C- and J-shocks driven into molecular gas through supersonic turbulence originating in the large-scale collision of NGC 7318b with gas in the group.     

In the SQ shock, powerful emission was also detected from [\ion{C}{2}] \citep{appleton2013}. The authors concluded that the emission was excited by collisions with the same warm H$_2$ that gave rise to the rotational H$_2$ lines \citep[see also][]{lesaffre2013}. This energy ultimately comes from the kinetic energy of the galaxy colliding with the intergroup medium, and cascades down to small spatial scales. As in Stephan's Quintet, the warm \Ht in the Taffy bridge appears to result from shock-heating of the molecular gas. Unlike SQ, where the intruder galaxy continues to provide a source of kinetic energy to drive the turbulent dissipation, the Taffy bridge may be in a decaying turbulence phase where energy from the recent collision has been largely expended. This may explain the large amount of neutral and molecular gas in the bridge. This study was motivated by the possibility that far-IR fine-structure lines, such as [\ion{C}{2}] and [\ion{O}{1}], may carry additional information about the cooling of gas in the Taffy bridge. 

The [\ion{C}{2}]157.7~$\micron$ transition can be excited by a variety of mechanisms, most commonly in PDRs surrounding young stars. In these regions, the [\ion{C}{2}] emission results from collisional excitation with warm H$_2$ heated by photoelectrons ejected from small grains and PAHs by UV radiation \citep{watson1972, glassgold1974, draine1978, tielens1985, hollenbach1989, bakes1994}. [\ion{C}{2}] emission can extend into CO-faint molecular gas \citep[so-called ``dark molecules'';][]{wolfire2010}, especially in star forming regions in low metallicity galaxies \citep[e.g.,][]{cormier2010}. 

However, as mentioned earlier, \citet{peterson2012} argued that the warm H$_2$ emission was inconsistent with photoelectric heating. Other potential collisional partners with carbon include neutral hydrogen atoms or free electrons associated with ionized gas in \ion{H}{2} regions \citep{Goldsmith2012}. If there is sufficient energy density, cosmic rays and X-rays can also play a role. Thus, in the absence of strong star formation, the interpretation of \cii emission is not straightforward, especially in a region like the Taffy bridge where many of these ingredients are present. 

While there is little evidence of widespread star formation in the Taffy bridge, there is an extragalactic \ion{H}{2} region (hereafter X-\ion{H}{2}) to the southwest of UGC 12915 (Figure~\ref{fig:regions}). The region is faint in broad-band images, but more prominent in H$\alpha$ \citep{bushouse1987} and Pa$\alpha$ \citep{komugi2012}. It lies in the region with the brightest CO emission, and is also near the position of the most luminous ultra-luminous X-ray source (ULX) in the Taffy system. This supports the idea that the star formation is occurring in a young stellar association embedded in the bridge \citep[see discussion in][]{appleton2015}. 

Given the previous evidence of a significant luminosity in the mid-IR H$_2$ lines, we explore the properties of the far-IR/sub-mm emission lines from the galaxies and bridge. We obtained \emph{Herschel} PACS observations in both the [\ion{C}{2}]158~$\micron$ and [\ion{O}{1}]63~$\micron$ emission lines, and a SPIRE FTS observation of a single pointing centered on the bridge.  In $\S$\ref{sec:observations}, we describe the details of the {\it Herschel} PACS and SPIRE observations, and the broad-band archival photometric observations. In $\S$\ref{sec:results} we present the results, including contour maps and line profiles of the [\ion{C}{2}] and [\ion{O}{1}] lines, as well as details of the SPIRE line detections and upper limits. We also compare the results with previously published CO~(1--0) and \ion{H}{1} data. We analyze these results in $\S$\ref{sec:analysis}, including the modeling of spectral energy distributions derived from from archival data, including {\it Herschel} photometry. We summarize our findings in $\S$\ref{sec:summary}. 
We assume a distance to the galaxies of 62~Mpc based on a  mean heliocentric
velocity for the system of 4350~km~s$^{-1}$, and a Hubble constant of 70~km~s$^{-1}$~Mpc$^{-1}$.
\\
\section{Observations} \label{sec:observations} 

\subsection{Herschel-PACS Spectroscopy}
Spectroscopic observations of the Taffy were made using the Photodetector Array Camera and Spectrometer (PACS; \citealt{Poglitsch2010}) integral field unit (IFU) on board the \emph{Herschel Space Observatory} \citep{pilbratt2010}\footnote{Herschel is an ESA space observatory with science instruments provided by European-led Principal Investigator consortia and with important participation from NASA.}. The spectra were obtained on 2012 January 30 as part of an open time program (program ID: OT1\_pappleto\_1; PI: Appleton: obsids = 1342238415/6). The PACS IFU uses a $5\times5$ grid of $9\farcs4 \times 9\farcs 4$ spatial pixels to obtain a $47\arcsec \times 47\arcsec$ field of view. Observations of the [\ion{C}{2}]157.74~$\micron$ and [\ion{O}{1}]63.18~$\micron$ lines were made using a $3 \times 3$ raster map (stepsize = 38 arcsec and repeated 4 and 5 times respectively) in chop-nod mode sampling a region of 1.83 $\times$ 1.83 arcmin$^2$, which included both galaxies and the bridge. The angular resolution in the [\ion{C}{2}] and [\ion{O}{1}] lines was (FWHM) 9\farcs4 and 3.\arcsec8 respectively, sampled onto $9\farcs4 \times 9\farcs4$  pixels. Total integration times per pointing were 23.5 and 50.0 minutes, with a total execution time for the full map of 3.9 and 8.2 hrs for the [\ion{C}{2}] and [\ion{O}{1}] observations, respectively\footnote{Observations centered on the bridge were also attempted of the OH $^2$$\Pi$$_{3/2}$ 5/2-3/2 $\lambda$-doublet line transitions (119.9/120.17$\micron$, obsid = 1342213131), but no emission was detected.}.   

The data were reduced using the Herschel Interactive Processing Environment (HIPE) version 15.0, and individual spectral-cube pointings were combined and regridded onto a sky projection with a pixel scale of $3\arcsec$. The spectral resolution in the \cii and \oi lines is 235~km~s$^{-1}$ and 85~km~s$^{-1}$ respectively. The spectra, observed in ``range mode'' to ensure coverage of the broad velocity field, span wavelengths from 159.4--160.8~$\micron$ (2570~km~s$^{-1}$) and 64.0--64.4~$\micron$ (1965 km~s$^{-1}$) respectively for the the [\ion{C}{2}] and [\ion{O}{1}] lines, centered on a redshift of $z = 0.0145$.  

Spectra were extracted from the final data cubes from regions covering both galaxies and the bridge, as shown in Figure~\ref{fig:regions}. The four bridge regions were chosen to provide nearly complete coverage of the bridge, and to provide information about how the gas properties vary across this part of the system. Their positions are listed in Table~\ref{tab:coords}. In addition, extraction regions were selected to sample X-\ion{H}{2}, the extended emission from the northwest part of the northern galaxy (UGC~12915), and the nucleus of the southern galaxy (UGC~12914). The nuclear region of the northern galaxy would be contaminated by extended emission from the disk, so no attempt was was made to extract a nuclear spectrum from this galaxy.


\subsection{Herschel SPIRE Spectroscopy}
The Taffy galaxies were observed with the SPIRE  Fourier Transform Spectrometer \citep{Griffin2010} on OD 1125 (program ID: OT1\_pappleto\_1; PI: Appleton: obsids = 1342246980) with a total integration time of 3.7~hrs on-source. The FTS has two detector arrays, the Spectrometer Long Wave (SLW) and Spectrometer Short Wave (SSW), which covered the long and short wavelength ranges, with a small overlap in wavelength. The complete spectrum covers the frequency range 455--1600~GHz ($\sim$194--671$\micron$). The FWHM of the SPIRE beam is 35 arcsec at 809.3~GHz, the rest frequency of the \ion{C}{1} $^3$P$_2$$\rightarrow$$^3$P$_1$ line. The observations were made with a single pointing in ``sparse mode'' (no mapping) with a resolving power ranging between 372 $< R <$ 1288 over the full range of the spectra. The central beam was pointed at the bridge center at $\alpha$(J2000) = 00$^{\rm h}$01$^{\rm m}$40$\fs$3, $\delta$(J2000) = 23$\degr$29$\arcmin$22$\farcs$0. Although other beams, apart from the central one, intersect with parts of the two galaxies, the emphasis in the paper is on the bridge, and so the other spectra will not be described in the current paper.    

SPIRE data reduction was performed using HIPE 15. The observed lines were extracted using a custom routine, after subtracting the appropriate background which was observed on that date and at the same resolution. This routine implemented a Sinc function to match the line profiles.

\subsection{Multi-wavelength Broad-band Photometry}\label{sec:sed}
Broad-band multi-wavelength coverage of the Taffy system was used to construct Spectral Energy Distributions (SEDs) for several extraction regions (see later discussion). These data were extracted from various archives, and included UV coverage from \emph{Galaxy Evolution Explorer} (\emph{GALEX}; \citealt{martin2005}), visible wavelengths from the Sloan Digital Sky Survey DR 7 (SDSS\footnote{Funding for SDSS-III has been provided by the Alfred P. Sloan Foundation, the Participating Institutions, the National Science Foundation, and the U.S. Department of Energy Office of Science. The SDSS-III web site is http://www.sdss3.org/.}), mid-IR from \emph{Spitzer} IRAC and MIPS \citep[][from NASA's Infrared Science Archive -- IRSA\footnote{This research has made use of the NASA/IPAC Infrared Science Archive, which is operated by the Jet Propulsion Laboratory, California Institute of Technology, under contract with the National Aeronautics and Space Administration.}]{werner2004, rieke2004, fazio2004} , and far-IR PACS photometry from the \emph{Herschel} Science Archive\footnote{http://archives.esac.esa.int/hsa/whsa/ and IRSA}. \\

To estimate the physical properties of the galaxies we have used MAGPHYS \citep{daCunha2008,daCunha2010} to fit the observed SEDs with sets of model templates. This code accounts for the global energy balance between the optical and the infrared and uses a Bayesian approach which draws from a large library of random models encompassing all plausible parameter combinations to derive the probability distribution function of various galaxy parameters, such as the stellar masses, star formation rates (SFRs), and both the total- and far-infrared luminosity for each of the regions. These values are listed in Table~\ref{tab:sed}. \\


\section{Results} \label{sec:results}
\subsection{Spatial Distribution of the [\ion{C}{2}] and [\ion{O}{1}] emission}\label{sec:spatialDist} 

\begin{figure*}
\includegraphics[width=0.9\textwidth]{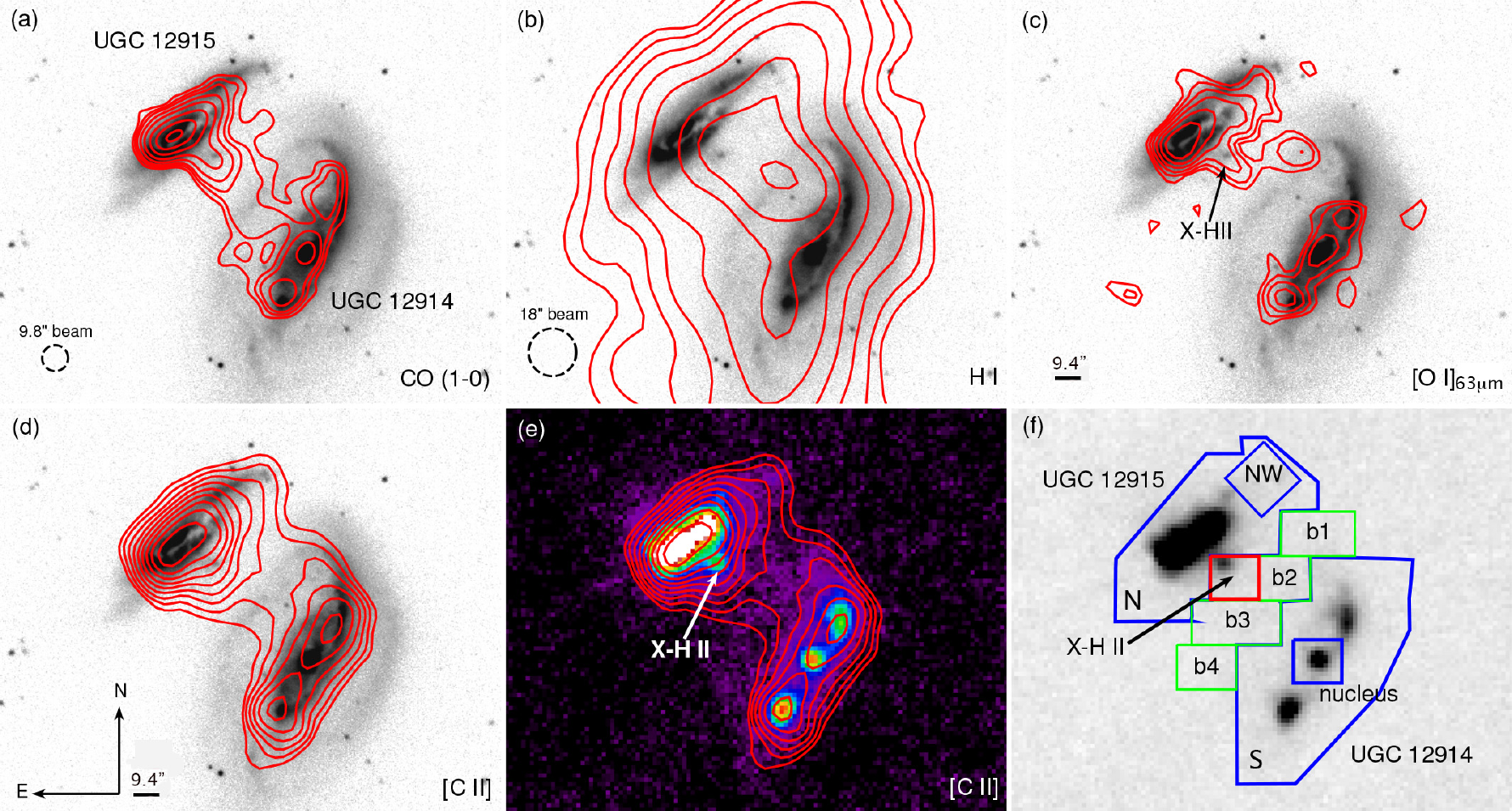}
\caption{(a) SDSS r-band image of Taffy system (logarithmic stretch), with red contours showing the integrated CO~(1--0) emission \citep{gao2003}. (b) Integrated \ion{H}{1} contours \citep{condon1993} on SDSS r-band image. (c) Integrated [\ion{O}{1}]63.2~$\micron$ contours based on our observations with the {\it Herschel} PACS spectrometer (this paper) on r-band (see Section~\ref{sec:contours}). (d) Integrated [\ion{C}{2}]157.7~$\micron$ contours using the {\it Herschel} PACS spectrometer (this paper) on r-band (see Section~\ref{sec:contours}). Images (c) and (d) are sampled onto 3 arcsec pixels from the PACS native pixel scale of 9.4 arcsecs. (e) Same contours of [\ion{C}{2}] as (d), but superimposed on a de-archived image from {\it Herschel} PACS photometer at $\lambda$70~$\micron$ displayed with a logarithmic stretch. The extragalactic \ion{H}{2} region (see text) is labeled as X-\ion{H}{2}. (f) Spectral extraction regions used for comparison of various spectra superimposed on a linear gray-scale version of the {\it Herschel} PACS~70~$\micron$ image. The offset of the X-\ion{H}{2} region box from the center of the optical counterpart, was chosen, based on inspection of the channel maps in the [\ion{C}{2}] data cube, to minimize contamination of emission from the disk of UGC 12915. Some contamination will be expected.} \label{fig:regions}
\end{figure*} 

To provide background to our observations, we present CO \citep{gao2003} and \ion{H}{1} \citep{condon1993} contours of the Taffy system superimposed on an optical image of the galaxies in Figure~\ref{fig:regions} (a) and (b). The CO interferometer observations, which have a similar spatial resolution to the PACS [\ion{C}{2}] observations, show a clear bridge extending from UGC 12915 to UGC 12914. The \ion{H}{1} observations, with much poorer spatial resolution, show a large concentration of \ion{H}{1} peaking between the galaxies. In addition to the difference in spatial resolution, the main centroid of the \ion{H}{1} emission may be offset towards the northwest compared with the CO emission.

Figure~\ref{fig:regions} (c) presents integrated emission contours of [\ion{O}{1}]63~$\micron$ derived from the PACS observations. The inner disks of both UGC 12915 and UGC 12914 are strongly detected in [\ion{O}{1}]. A ridge of emission in UGC 12915 follows the dark dust lanes in the disk, and extends towards the northwest where the emission becomes more extended. In UGC 12914, the emission is concentrated towards the nucleus, and in two bright regions further out in the disk. Clumpy emission from [\ion{O}{1}] is present between the galaxies, including the X-\ion{H}{2} region, and a prominent clump between the galaxies. Fainter emission is present in the bridge, but this is not well represented in the integrated map. \\

Figure~\ref{fig:regions} (d) and (e) show the distribution of [\ion{C}{2}] emission superimposed on both the optical image and the 70~$\micron$ PACS dust continuum image obtained from the {\it Herschel} Science Archive. The [\ion{C}{2}] distribution is similar to that of the [\ion{O}{1}], though there are some differences. For example, the [\ion{C}{2}] emission appears to be strong relative to the [\ion{O}{1}] in the northwest part of UGC 12915. Furthermore, the nucleus of UGC 12914 seems more prominent in the [\ion{O}{1}] emission, as compared with [\ion{C}{2}]. The [\ion{C}{2}] is also well-correlated with the faint dust emission, including faint emission in the bridge. It is notable that the southern boundary of UGC 12914 is very sharply defined, similar to the CO and radio continuum distributions (the latter not shown here but see \citealt{condon1993}). The emission does not extend into the outer southern ring of UGC 12914, indicating that the ISM of this galaxy has been severely stripped from the disk by the collision with UGC 12915. 

Based on the gas distribution, several spectral extraction boxes were defined, as shown in Figure~\ref{fig:regions} (f). The large polygons  are intended to capture all of the emission from the galaxies. The bridge is broken up into four regions b1~--~b4 to allow us to look for variations in the gas properties of the bridge. We also extracted a spectrum from the unusual extragalactic X-\ion{H}{2} region, and the nucleus of UGC 12914. The UGC 12915 nucleus could not be isolated in either the [\ion{C}{2}] or 70~$\micron$ emission and was not extracted. Instead, we extracted the extended emission in the northwest part of UGC 12915. This is a region where the underlying far-IR continuum is weak, but the [\ion{C}{2}] emission is quite pronounced. \\

\subsection{Velocity Distribution of [{\ion{C}{2}]} and [{\ion{O}{1}}] emission} \label{sec:contours} 
\begin{figure*}
\includegraphics[width=\textwidth]{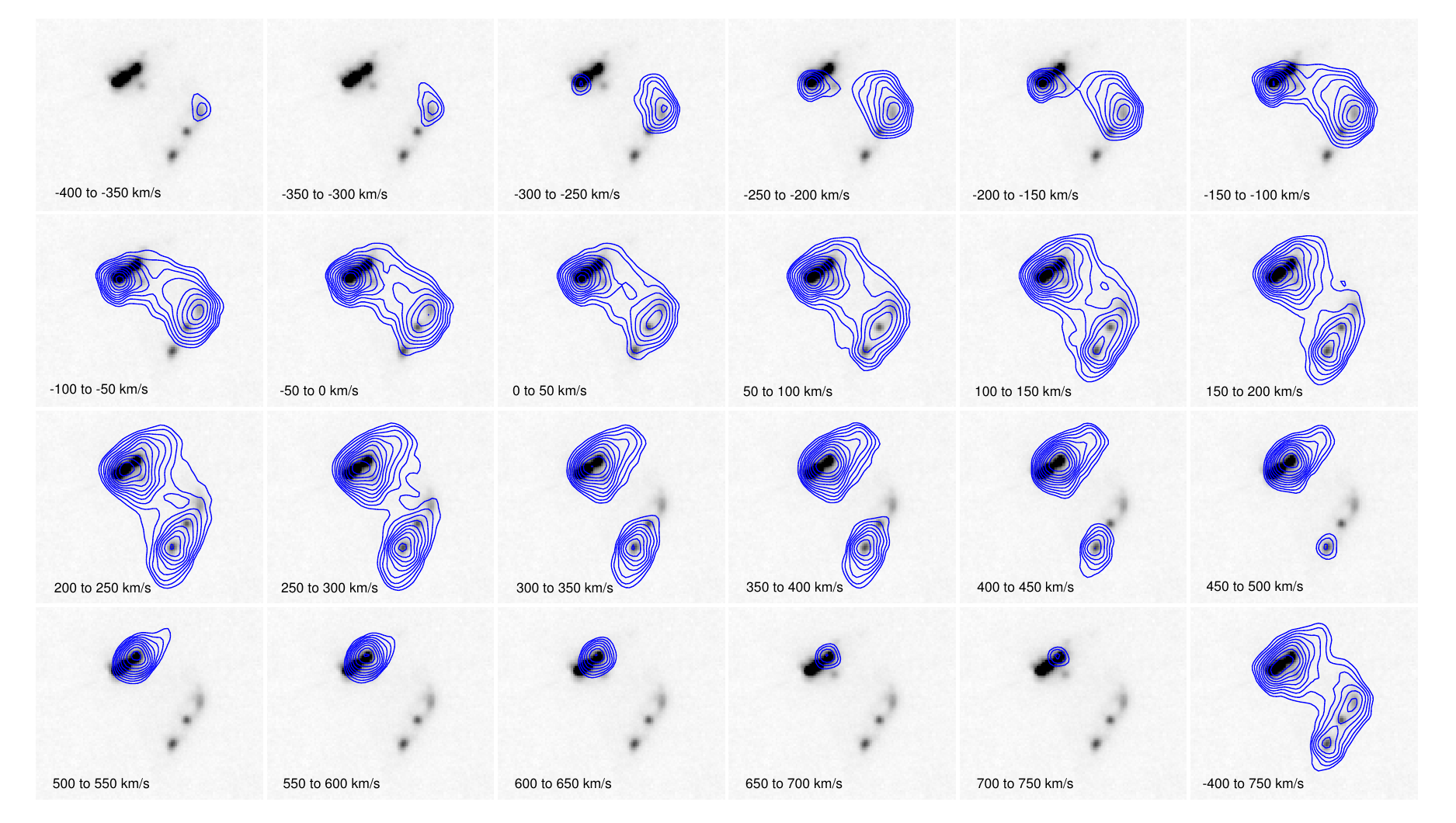}
\caption{[\ion{C}{2}] velocity contours over PACS 70~$\micron$ image. The contour levels begin at 3.0~Jy~km~s$^{-1}$ and increase successively by a factor $\sqrt{2}$. The last panel is integrated over the entire velocity range with contour levels beginning at 20~Jy~km~s$^{-1}$ and increase successively by a factor of $\sqrt{2}$. Zero velocity is equal to 4350~km~s$^{-1}$ heliocentric.} \label{fig:ciiMaps}
\end{figure*}

Figure~\ref{fig:ciiMaps} shows velocity contour maps for the [\ion{C}{2}] line. As previously noted by many authors, the two galaxies appear to be counter-rotating. The emission in UGC~12914 is from $v\sim-400$ to 500~km~s$^{-1}$, with negative velocities at the northern end of the galaxy and positive toward the south\footnote{All Velocities are heliocentric and relative to a radial velocity of 4350~km~s$^{-1}$}. The emission at the southern end of the galaxy extends along the visible spiral arm farther than is apparent from the integrated contours of Figure~\ref{fig:regions}, with extended emission from $v\sim$200 to 400~km~s$^{-1}$. On the other hand, UGC~12915 is detected from $v\sim-300$ to 750~km~s$^{-1}$ with negative velocities toward the south and positive toward the north: the opposite of UGC~12914. The emission peak is confined to the same compact area as the 70~$\micron$ emission at all velocities. The extended emission in the northwest part of the galaxy is evident from $v\sim150$ to 450~km~s$^{-1}$. This is a wider velocity range than the emission along the southern arm of UGC 12914.\\

Emission in the bridge is detected from $-300$ to $350$~km~s$^{-1}$, starting just south of the northern arm of UGC~12914 and moving across the bridge toward the center of UGC~12915 at increasing velocities. The X-\ion{H}{2} region near UGC~12915 appears at velocities from $v\sim -100$ to $300$~km~s$^{-1}$. While much of the emission from X-\ion{H}{2} appears at the same velocities as the diffuse bridge emission, it is also strong from 200 to 300~km~s$^{-1}$, where the diffuse bridge emission has largely faded.  \\

\begin{figure*}
\includegraphics[width=\textwidth]{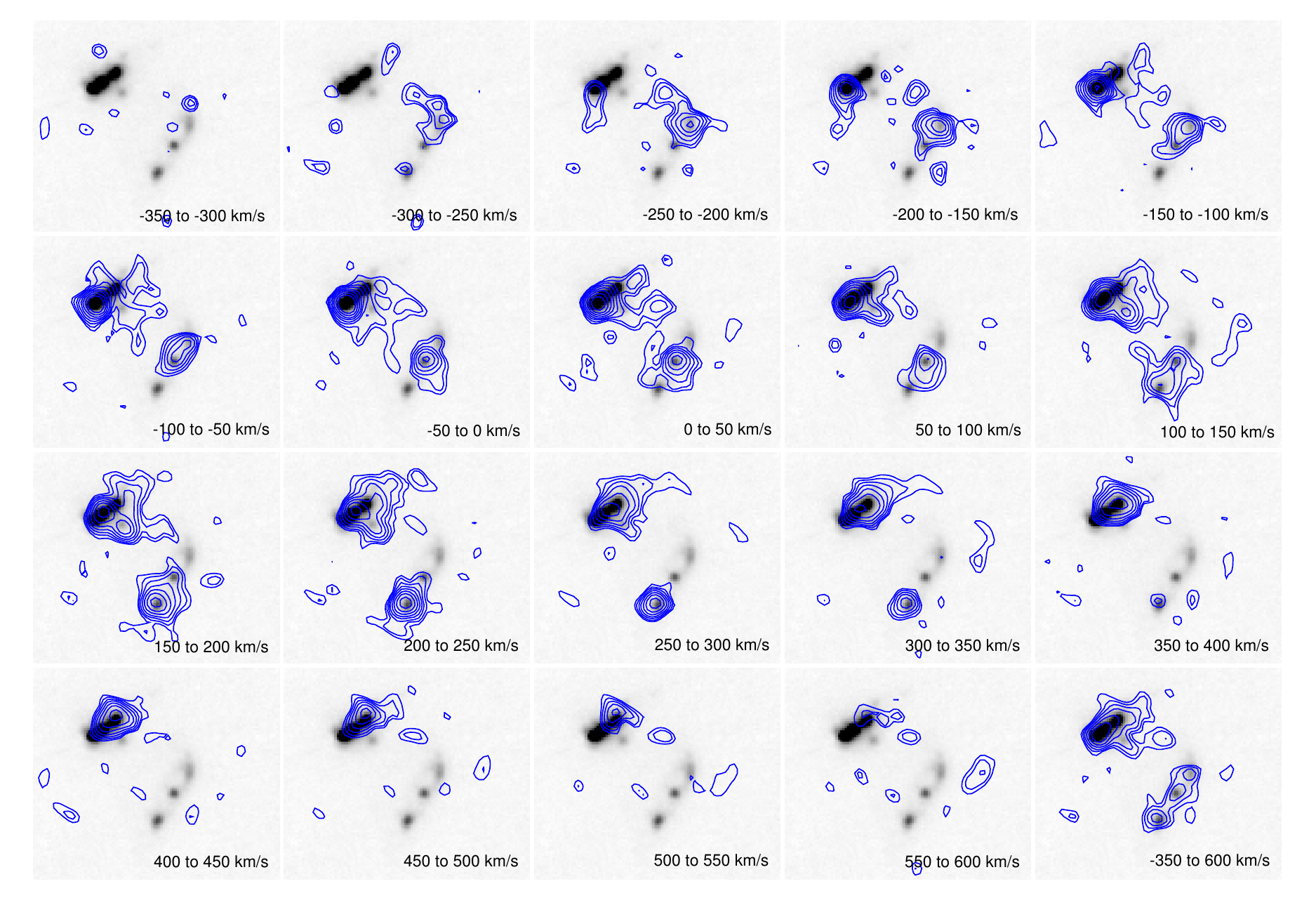}
\caption{[\ion{O}{1}] velocity maps over PACS 70~$\micron$ image. The contour levels begin at 1.0~Jy~km~s$^{-1}$ and increase successively by a factor of $\sqrt{2}$. The last panel is integrated over the entire velocity range with contour levels beginning at 10~Jy~km~s$^{-1}$ and increase successively by a factor of $\sqrt{2}$. Zero velocity is equal to 4350~km~s$^{-1}$ heliocentric. } \label{fig:oiMaps}
\end{figure*}

\oi velocity contours are presented in Figure~\ref{fig:oiMaps}. UGC 12915 is detected from $v = -250$ to 600~km~s$^{-1}$ and, as in [\ion{C}{2}], shows extended emission to the northwest. Clumpy extended emission appears to extend along the northwest disk of UGC 12915, and extend into the bridge from $\sim$0 to 200~km~s$^{-1}$. Starting at $\sim$200~km~s$^{-1}$, and continuing through $\sim$350~km~s$^{-1}$, the emission extends well past the northern arm. The X-\ion{H}{2} region is better defined in the [\ion{O}{1}] maps than in [\ion{C}{2}]. The narrow northwestern arm of UGC 12914 extends out of the disk and into the bridge at velocities of $-300$ to $-200$~km~s$^{-1}$. \\

\subsection{A comparison between the  [\ion{C}{2}] and [\ion{O}{1}] spectra, and that of CO (1--0) and HI} 
\begin{figure}
\includegraphics[width=0.47\textwidth]
{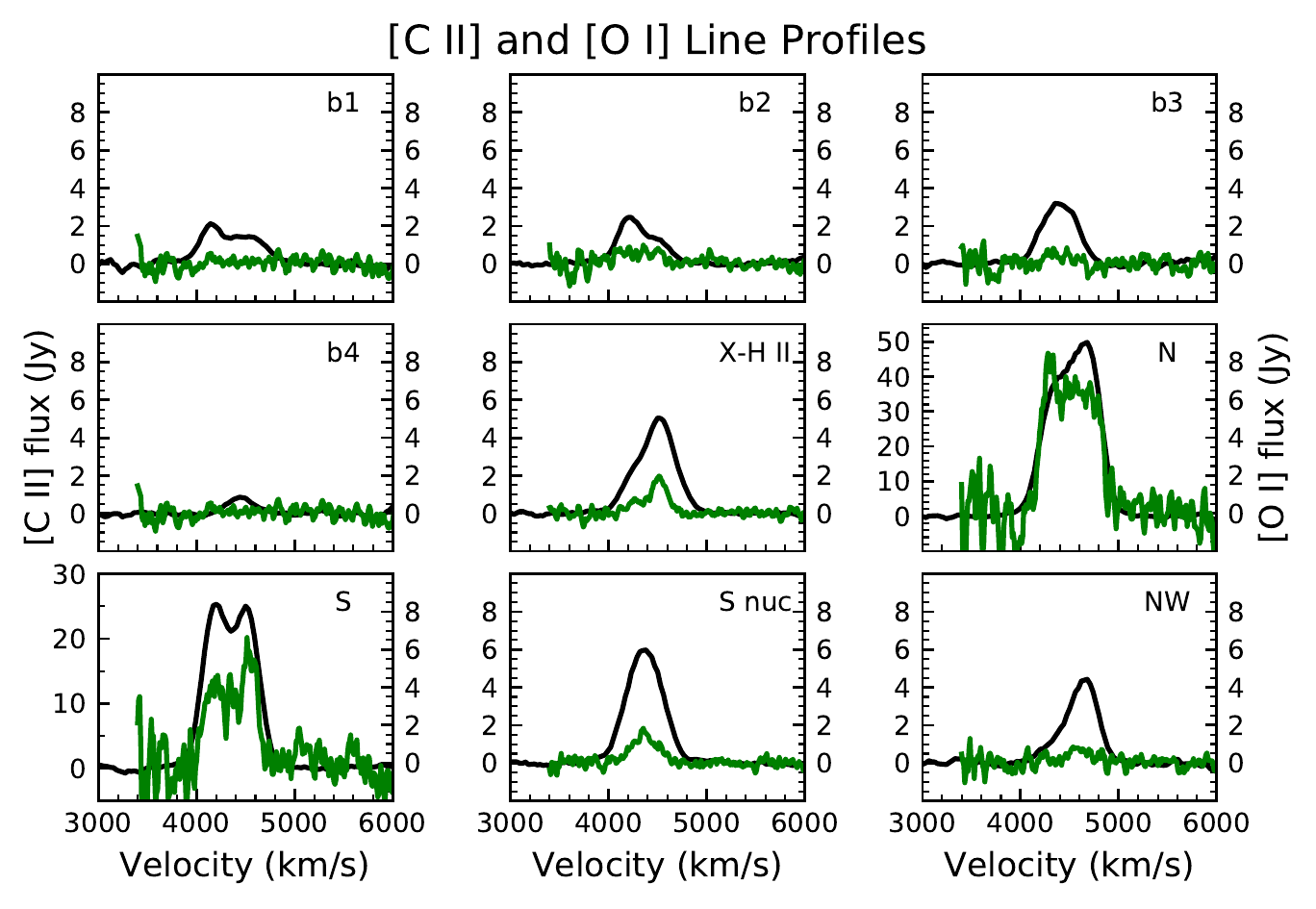}
\caption{[\ion{C}{2}]157.7$\micron$ line profiles (black) and [\ion{O}{1}]63.1~$\micron$ (green) with spectral resolutions of 235~km~s$^{-1}$ and 85~km~s$^{-1}$ respectively. The extraction regions are shown in Figure~\ref{fig:regions} and their positions and sizes are listed in Table~\ref{tab:coords}.} \label{fig:profiles_cii}
\end{figure}

Figure~\ref{fig:profiles_cii} shows the profiles of the [\ion{C}{2}] (black lines) and  [\ion{O}{1}] (green lines) lines for each of the extraction regions. The [\ion{C}{2}] is strong (peak flux densities of 50 and 25~Jy integrated over the two galaxies; labeled N and S for UGC 12915 and UGC 12914 respectively), and weaker in the bridge and the X-\ion{H}{2} region (in the peak flux range of a few Jy; labeled b1~--~b4 and X-\ion{H}{2}). The bridge and X-\ion{H}{2} extractions show very broad line profiles that suggest multiple components. Similar, though weaker and noisier, broad profiles are also seen in the [\ion{O}{1}] extractions (green lines) shown in Figure~\ref{fig:profiles_cii}. The [\ion{O}{1}]63~$\micron$ observations have higher resolving power (85~km~s$^{-1}$) and are better able to resolve the complex velocity structure of the bridge. For example, the X-\ion{H}{2} region spectrum in [\ion{O}{1}] shows a narrower high-velocity feature and a broader low-velocity feature, consistent with the lower resolution [\ion{C}{2}] profile. The [\ion{O}{1}] emission is weakly detected at positions b2 and b3, is marginally detected in b1, and is not detected in b4. As in the case of the [\ion{C}{2}] profile, the [\ion{O}{1}] emission at bridge positions b1 and b2 shows multiple components, whereas b3 is better represented by a single component. 

The line fluxes and velocity components for the [\ion{C}{2}] and [\ion{O}{1}] lines are given in Tables~\ref{tab:cii} and \ref{tab:oi}, respectively, with line ratios provided in Table~\ref{tab:linerats}. Line fluxes were measured using the \emph{ISO} Spectral Analysis Package (ISAP; \citealt{sturm1998}). Each line was measured using either a single or double line fit to these spectra. The two components usually provided a better match to [\ion{C}{2}] profiles. There were, however, cases where the data were fit with a single component, such as b2 and b3, where the [\ion{O}{1}] is weak. In b1 and b4, the [\ion{O}{1}] emission is sufficiently weak that we are only able to provide 3$\sigma$ upper limits. \\

We note that in the bridge position b3, the observed line ratio of [\ion{C}{2}]/[\ion{O}{1}] = $2.75 \pm 0.66$, which is similar to that seen in several of in the intergalactic shocked regions in the Stephan's Quintet system \citep{appleton2013}, and with similar [\ion{C}{2}] surface intensity. By analogy with that work, we estimate similar derived properties to that gas ($T \sim 200$~K) diffuse ($10^3$~cm$^{-3}$) gas in the Taffy bridge (see Figures 6 and 7 of that paper). \\

The emission in [\ion{C}{2}] from the NW-disk extraction region in UGC 12915 is quite strong and broad, given its weak far-IR emission (see Figure 1f), and is much weaker in [\ion{O}{1}]. It is possible that this emission is bridge material superimposed on the northern disk, as it shares similar properties to other material in the bridge. X-ray emission, which we suggested was shock-heated gas left over from the initial collision between the Taffy galaxies \citep{appleton2015}, also occupies this region of the north-west disk of UGC 12915 and the bridge. 

Both the [\ion{C}{2}] and [\ion{O}{1}] profiles of the nucleus of UGC 12914 (labeled S-nuc in Figure~\ref{fig:profiles_cii}) show broad lines (FWHM 354 and 288~km~s$^{-1}$ respectively for [\ion{C}{2}] and [\ion{O}{1}] after deconvolving with the instrument profile). The {\it Chandra} X-ray spectral index measurements of \citet{appleton2015} showed possible evidence for a weak low-luminosity AGN in this galaxy. \\

\begin{figure}
\includegraphics[width=0.5\textwidth]{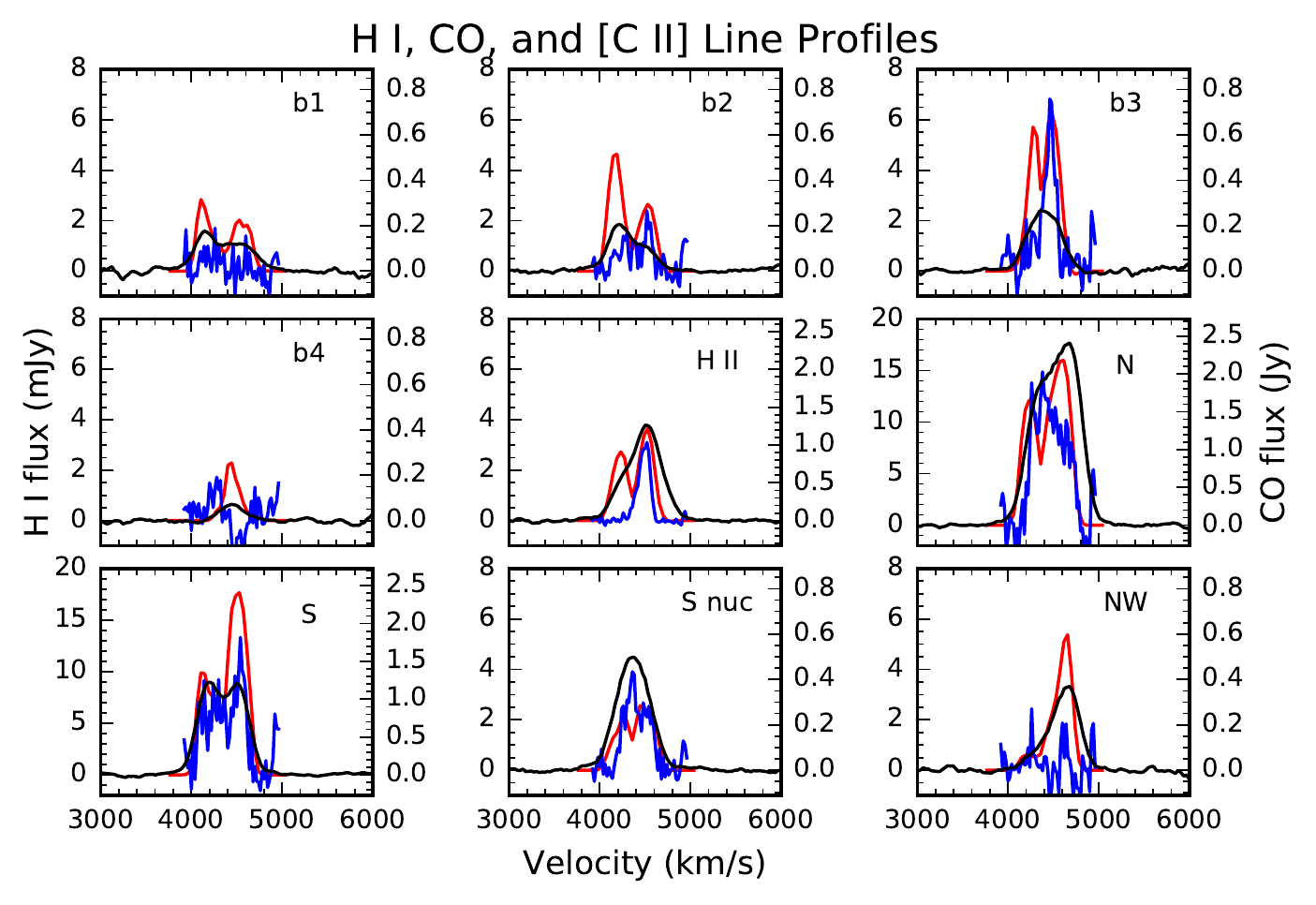}
\caption{Line profiles for \ion{H}{1} (red), CO (1--0) (blue), and [\ion{C}{2}] (black). The [\ion{C}{2}] units are arbitrary; for values, see Figure~\ref{fig:profiles_cii}. The extraction regions are shown in Figure~\ref{fig:regions} and their positions and sizes are listed in Table~\ref{tab:coords}. The \ion{H}{1} data have a spectral resolution of 42.4~km~s$^{-1}$, while the resolution in CO is 20~km~s$^{-1}$.}  \label{fig:profiles_hcocii}
\end{figure} 

\begin{figure}
\includegraphics[width=0.5\textwidth]{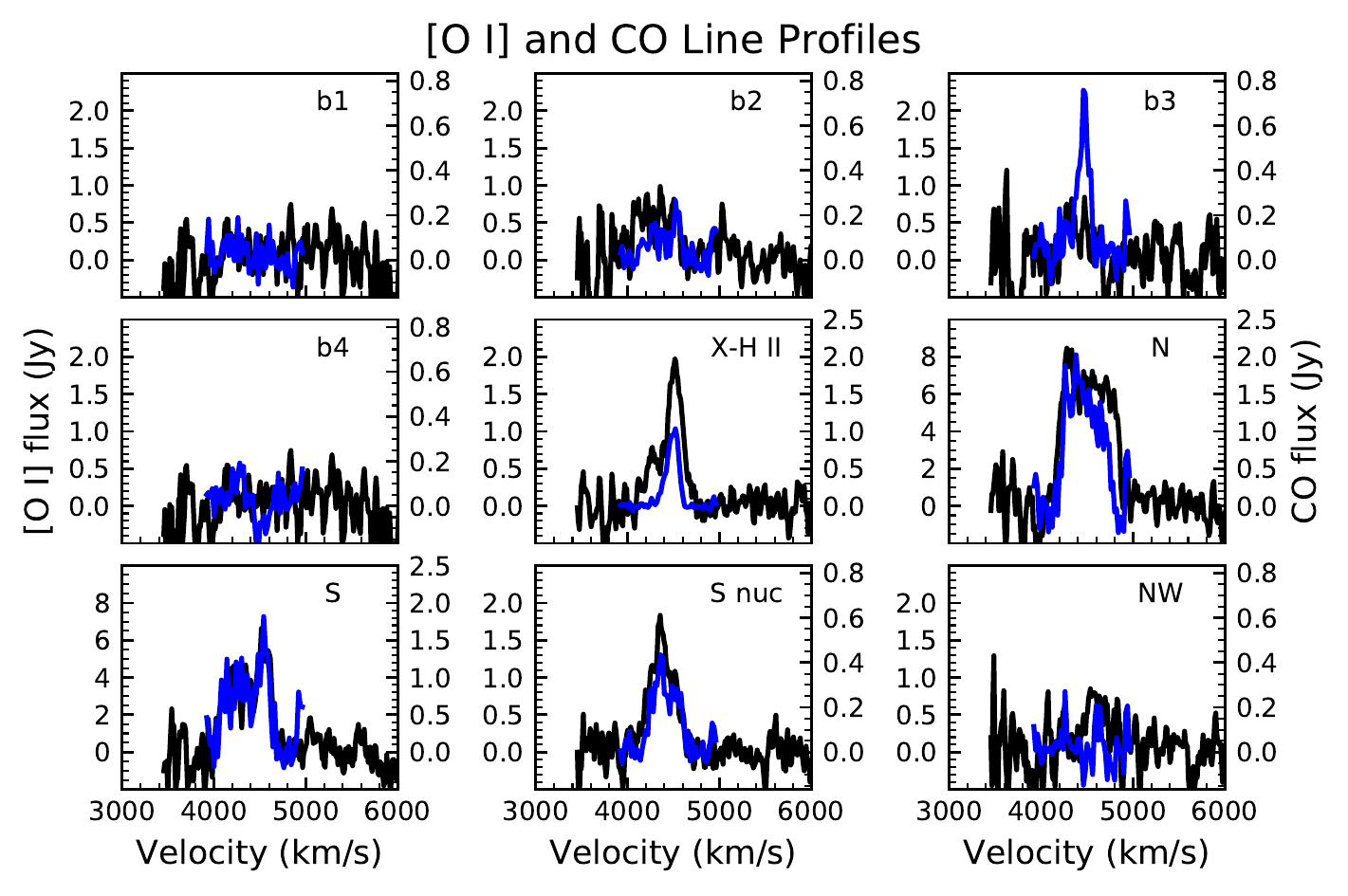}
\caption{Line profiles for [\ion{O}{1}] (black), CO (1--0) (blue). The extraction regions are shown in Figure~\ref{fig:regions} and their positions and sizes are listed in Table~\ref{tab:coords}.}  \label{fig:profiles_oico}
\end{figure} 





To compare the PACS spectra of the extracted regions with the  \citep{condon1993} and CO data from BIMA CO~(1--0) \citep{gao2003} we obtained calibrated data cubes of both sets of observations and performed extractions over similar areas to those of Figure~\ref{fig:regions}. Figure~\ref{fig:profiles_hcocii} shows the emission profiles of [\ion{C}{2}] with \Hi and CO~(1--0), and Figure~\ref{fig:profiles_oico} compares the [\ion{O}{1}] and CO~(1--0) line profiles. The CO~(1--0) spatial resolution (9.8$\arcsec \times$ 9.8$\arcsec$) is similar to that of the [\ion{C}{2}], but the \ion{H}{1} data has much poorer resolution (18$\arcsec \times$ 18$\arcsec$), and so this must be taken into account. All of the {\it Herschel} spectral extraction boxes are large enough to include at least one beam of the \ion{H}{1} VLA data. Nevertheless, these latter profiles are likely to be more affected by resolution effects than the other spectra. In some of our comparisons, we have convolved the CO and [\ion{C}{2}] data to the same resolution as the \ion{H}{1} data.   \\

There are several regions in the Taffy bridge where where the [\ion{C}{2}], \ion{H}{1}, and CO~(1--0) profiles differ, even after taking into account the differing resolutions. The biggest differences are between the CO and the \ion{H}{1} in regions b2, b3, and in X-\ion{H}{2}. The \ion{H}{1} is double-peaked, with a low- and high-velocity component that is similar to the [\ion{C}{2}] profiles, whereas the CO is single-peaked with only the high-velocity component being dominant. This is particularly striking in the X-\ion{H}{2} spectrum, where the CO~(1--0) profile shows almost no emission in the low-velocity component, whereas the \ion{H}{1} and [\ion{C}{2}] profiles exhibit both components. In region X-\ion{H}{2}, the [\ion{O}{1}] emission is more strongly peaked on the high-velocity component, corresponding with the CO, but also shows weaker emission in the low-velocity component of the double-profile. In other regions of the bridge, the situation is less clear-cut. For example, in b2 and b3, the CO contains a bright high-velocity component, but a weaker low-velocity component, both of which are represented in [\ion{C}{2}] and weakly in [\ion{O}{1}]. 


The fact that the [\ion{C}{2}] and \ion{H}{1} share similar double-peaked line-shapes in parts of the bridge, but that the CO~(1--0)-emitting gas primarily dominates the higher-velocity peak, shows that the two kinematic components contain a different mix of \ion{H}{1} and CO-bright H$_2$ \citep[we cannot, of course, rule out CO-faint H$_2$;][]{wolfire2010}. Both velocity components also contain an unknown (but potentially large) fraction of mid-IR-emitting warm molecular hydrogen \citep[the {\it Spitzer} IRS observations did not have sufficient resolution to separate the high and low velocity components;][]{peterson2012}. It is therefore very likely that both neutral and molecular hydrogen contribute to the collisional excitation of [\ion{C}{2}]157~$\micron$ line in the presence of a small degree of UV ionization. We will return to this point in $\S$\ref{sec:analysis}.   \\

\subsection{The SPIRE spectrum of the Bridge Center}

\begin{figure}
\plotone{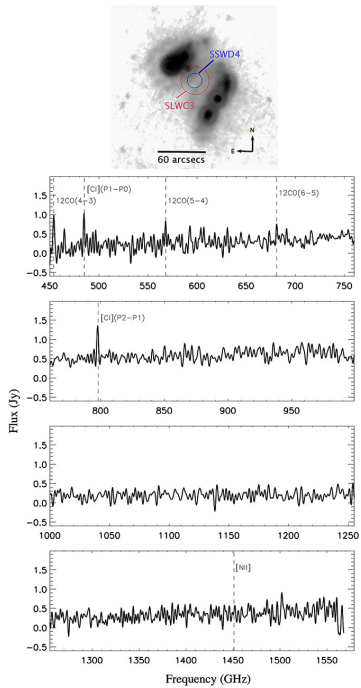}
\caption{Top: 24~$\micron$ image of Taffy with the beams for the SWS and SLW detector pointings shown. First three spectral panels: The SPIRE FTS spectrum from detector SLW C3, centered on the Taffy bridge, divided into three parts for clarity, showing the detection of $^{12}$CO(6--5), $^{12}$CO(~5--4), [\ion{C}{1}]~(1--0) and [\ion{C}{1}]~(2--1). The identification of the $^{12}$CO~(4--3) line is close to the band-edge and should be treated with caution. The bottom panel shows part of the spectrum from the SSW-D4 detector (co-spatial with SLW C3) spectrum which includes [\ion{N}{2}]205~$\micron$, which is not detected. Extracted line fluxes and upper limits are given in Table \ref{tab:spire}.   \label{fig:spire}}
\end{figure}

Figure \ref{fig:spire} shows the spectrum of the bridge split into several pieces to zoom in on the relevant detected lines. Several lines are detected including the neutral carbon lines [\ion{C}{1}] $^3$P$_2$$\rightarrow$$^3$P$_1$ (rest frequency 809.3~GHz = 370.4~$\micron$), [\ion{C}{1}] $^3$P$_1$$\rightarrow$$^3$P$_0$ (rest frequency 492.2~GHz = 609.1~$\micron$; hereafter [\ion{C}{1}]~(2--1) and [\ion{C}{1}]~(1--0) respectively). The $^{12}$CO~(6--5) and $^{12}$CO~(5--4) are weakly detected at a lower level of significance. The $^{12}$CO~(4--3) line appears to be stronger; however, it is very close to the band edge so its existence must be treated with caution. We also measured a 3$\sigma$ upper limit for the [\ion{N}{2}]205~$\micron$ line. The line measurements are presented in Table~\ref{tab:spire}. 

The detection of two [\ion{C}{1}] lines is perhaps not surprising given that they likely trace diffuse molecular gas with similar density to that of the CO~(1--0) emitting gas \citep{papadopoulos2004}. The critical densities of the 2--1 and 1--0 lines of [\ion{C}{1}] are 500 and 1000~cm$^{-3}$ respectively, which is very similar to that of CO~(1--0) (440~cm$^{-3}$). Interestingly, the ratio of the 2-1 and 1-0 lines, $R$$_{[{\rm C I}]}$ = 0.7$\pm$0.1, is an outlier when compared with samples of normal and luminous infrared galaxies, where typical galaxies show values in the range 0.1 $<$ $R$$_{[{\rm C I}]}$ $<$ 0.5, and  is similar to that of NGC 6240 \citep{jiao2017}. This latter system's value of $R$$_{[{\rm C I}]}$ (= 0.81$\pm$0.09) is unusual, even compared with most (U)LIRGs, and is believed to be primarily shock-heated. We can quantify this if the [\ion{C}{1}] lines are optically thin. In that case the excitation temperature of the gas can be estimated from \citet{Walter2011}  to be T$_{ex}$ = 38.8 $\times$ [ln(2.11/$R$$_{[{\rm C I}]}$)]$^{-1}$~K , or 35~K. This relatively high excitation temperatures for [\ion{C}{1}], and presumably CO~(1--0) emitting molecular gas, is another indication of the unusual nature of the Taffy bridge, again suggesting shock-heating as a viable mechanism for heating the gas.  

Finally, extracting the [\ion{C}{2}]157.7~$\micron$ line flux from the same regions as the SPIRE SSW and SLW beams (5.4 and 22.0 $\times$ 10$^{-17}$~W~m$^2$ respectively), yields [\ion{C}{2}]/[\ion{C}{1}] = 11 (summing the fluxes from the [\ion{C}{1}]~(2--1) and 1--0 transitions) and  [\ion{C}{2}]/[\ion{N}{2}] $> 54$. The value of [\ion{C}{2}]/[\ion{C}{1}] is relatively low compared with values $\sim$35 for active star forming galaxies (derived from results shown in \citealt{Lu2015, Lu2017}). This is consistent with the low measured [\ion{C}{2}]/CO(1--0) ratios (see $\S$\ref{sec:covcii}), since the [\ion{C}{1}] and the CO~(1--0) sample similar phases. 

The non-detection of the [\ion{N}{2}]205~$\micron$ line, and a measured lower limit to the ratio of [\ion{C}{2}]/[\ion{N}{2}] $>54$ (where the [\ion{C}{2}] is extracted from the larger SLW beam), implies a small ionized gas fraction in the bridge. For pure ionized gas, the ratio [\ion{C}{2}]/[\ion{N}{2}]$\sim 4$ and is almost independent of density. In normal galaxies, [\ion{C}{2}] emission arises primarily from the neutral gas phase, with typical values of [\ion{C}{2}]/[\ion{N}{2}] lying between 11 and 22 (Croxall et al. 2017, based on KINGFISH galaxy sample). The lower limit of this ratio for the Taffy bridge implies that most of the [\ion{C}{2}] originates in a neutral or molecular phase. 

\subsubsection{CO~(1--0) versus [\ion{C}{2}] line fluxes in the Taffy} \label{sec:covcii}

CO~(1--0) line fluxes were extracted using the CO data cube of \citet{gao2003}; Table~\ref{tab:linerats} shows the [\ion{C}{2}]/CO ratios in the extraction regions. We find [\ion{C}{2}]/CO ratios of 465$\pm$47 and 670$\pm$67 for UGC 12914 and UGC 12915, respectively. These values are much lower than the mean value for starburst galaxies, and lower than the values $\sim$6300 found in galactic star formation regions by \citet{stacey1991}. They are also lower than the values typically found in normal galaxies $\sim$1500--3000, with large scatter. The [\ion{C}{2}]/CO ratios in the bridge are even lower, with b2 and b3 having ratios of $348\pm35$ and $218\pm22$, respectively. Low \ion{C}{2}]/CO values in normal spiral galaxies are often taken to imply that the galaxies are dominated by cool CO clouds with very little ambient UV radiation to excite [\ion{C}{2}] emission over the same regions. The low values of \ion{C}{2}]/CO(1--0) is consistent with our earlier result that the [\ion{C}{2}]/[\ion{C}{1}] ratio is also lower than that found in normal star forming galaxies. This supports the earlier idea that the [\ion{C}{1}] emission is probing the same gas phase as that of CO~(1--0).

\section{The origin of the [\ion{C}{2}] emission in the bridge} \label{sec:analysis}

\subsection{Star Formation in the Taffy System}

The star formation rates (SFRs) for the two galaxy extraction regions estimated using MAGPHYS SED fits are 2.77$\pm0.30$ and 1.13$\pm0.01$~$M_{\sun}$~yr$^{-1}$ for UGC~12915 and 12914, respectively. These values are in good agreement with those of \citet{appleton2015} who estimated the SFR using several different methods. 

The SFR can also be estimated using the [\ion{C}{2}] luminosities. Using the calibration of \citet{herrera-camus2015} with no IR color correction, we find SFRs of 5.2~$M_{\sun}$~yr$^{-1}$ for UGC 12915 and 2.6~$M_{\sun}$~yr$^{-1}$ for UGC 12914. For the two galaxies, the [\ion{C}{2}]-based estimate is higher than the SED-based estimate by a factor $\sim$2. However, the IR color temperatures of both galaxies are on the lower end of the distribution for normal galaxies, and this would act to reduce the overall SFR derived from this correlation, bringing them into closer agreement with the SED method.   \\

More interesting are the SFRs in the bridge and X-\ion{H}{2}. Based on the SED fit, we estimate a total bridge SFR of 0.13~$M_{\sun}$~yr$^{-1}$, excluding region X-\ion{H}{2}. This is in approximate agreement with estimates using the 24~$\micron$ emission from the same regions \citep{appleton2015}. On the other hand, if the [\ion{C}{2}] line luminosity is used to calculate the star formation rate, we estimate a much higher value of 0.64~$M_{\sun}$~yr$^{-1}$. Although there is a large spread in the SFR/[\ion{C}{2}] calibration, this excess implies that the [\ion{C}{2}] emission is boosted by processes other than normal star formation processes. 

\begin{figure}
\plotone{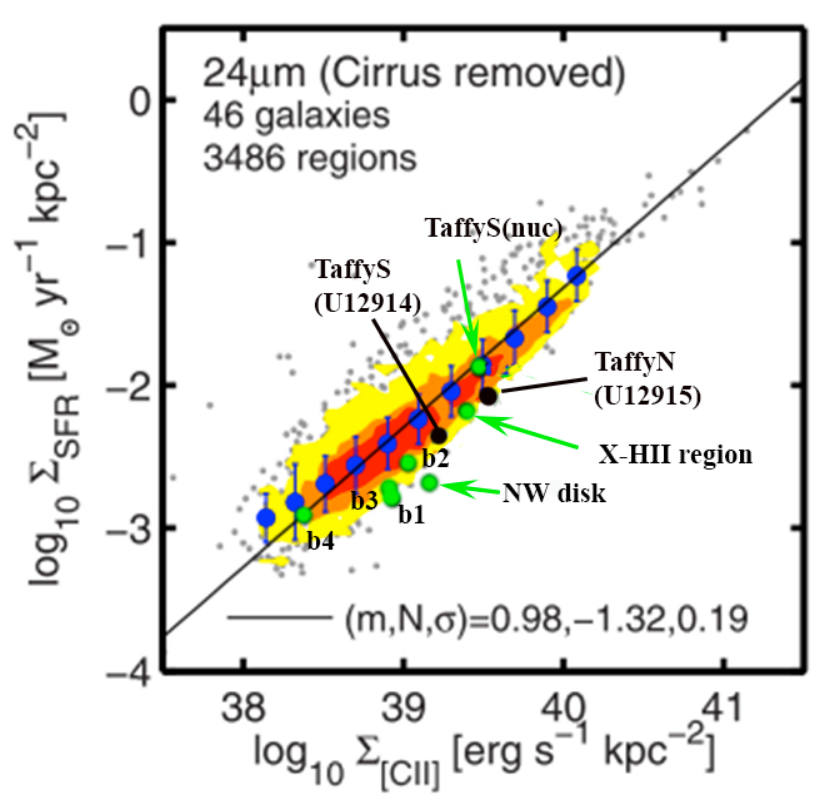}
\caption{Star formation rate surface density, estimated from 24~$\micron$ emission, vs. [\ion{C}{2}] surface brightness adapted from a figure by \citet{herrera-camus2015}. The colored regions show the density of points from galaxy regions studied by \citet{herrera-camus2015}. Black filled circles show the position on the same diagram of the Taffy galaxies, and the green filled circles show the bridge and other extracted regions from Figure~\ref{fig:regions}. Note that many of the Taffy regions lie on the extreme edge of the distribution for normal galaxies, suggesting that the [\ion{C}{2}] emission may be boosted there.}.\label{fig:sfr24um}
\end{figure} 

To explore this further, we attempt to quantify this excess in Figure~\ref{fig:sfr24um}, where we plot the SFR surface density $\Sigma_{SFR}$, determined from the {\it Spitzer} MIPS 24~$\micron$ image of the system, with the [\ion{C}{2}] surface brightness $\Sigma_{[C II]}$. The figure is taken from \citet{herrera-camus2015}, and includes a density distribution of points taken from a sample of 3486 regions in 46 normal galaxies. The two Taffy galaxies are marked as black circles in the diagram. The other extracted regions are marked in green. Except for the UGC 12914 nucleus, and region b4, most of the points, including the galaxies as a whole, lie on the extreme edge of the distribution for regions in normal galaxies. This makes the point that the [\ion{C}{2}] emission appears generally boosted, compared with that expected for normal star formation. Alternatively, Figure~\ref{fig:sfr24um} could be interpreted as showing lower star formation rate surface densities for a given [\ion{C}{2}] surface density, compared with normal galaxies. However, as we shall see, it is more likely that the [\ion{C}{2}] is boosted. We note that the region in the extreme northwest part of UGC 12915 (labeled NW disk in the figure) also shows unusually high [\ion{C}{2}] emission. In this respect it shows similarities with the bridge. Indeed, kinematically, it may be a projection of part of the northern bridge emission onto the galaxy (see Figures~\ref{fig:ciiMaps} and \ref{fig:oiMaps}). \\

\begin{figure}
\plotone{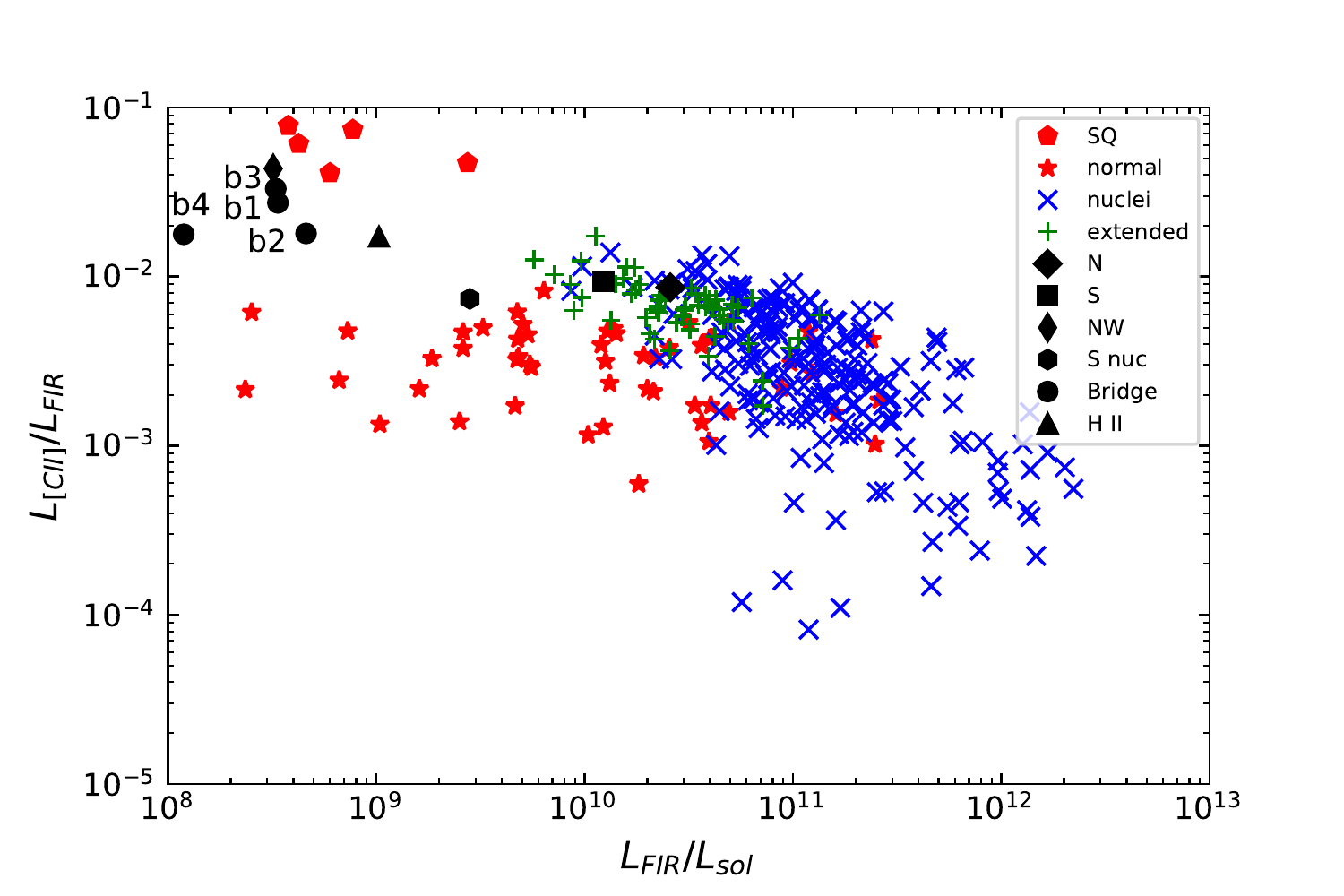}
\caption{[\ion{C}{2}]/FIR ratio for the bridge regions (black circles), extragalactic \ion{H}{2} region (black triangle), southern galaxy (black square), southern nucleus (black hexagon), and northern galaxy (black diamond). For comparison, we show SQ (red pentagons; \citealt{appleton2013}), normal galaxies from \citet{malhotra2001} (red stars), LIRG nuclei \citep{diazsantos2013} (blue crosses), and LIRG extended emission \citep{diazsantos2014} (green plus signs).} \label{fig:ciiToFIR}\textit{}
\end{figure} 
\begin{figure}
\plotone{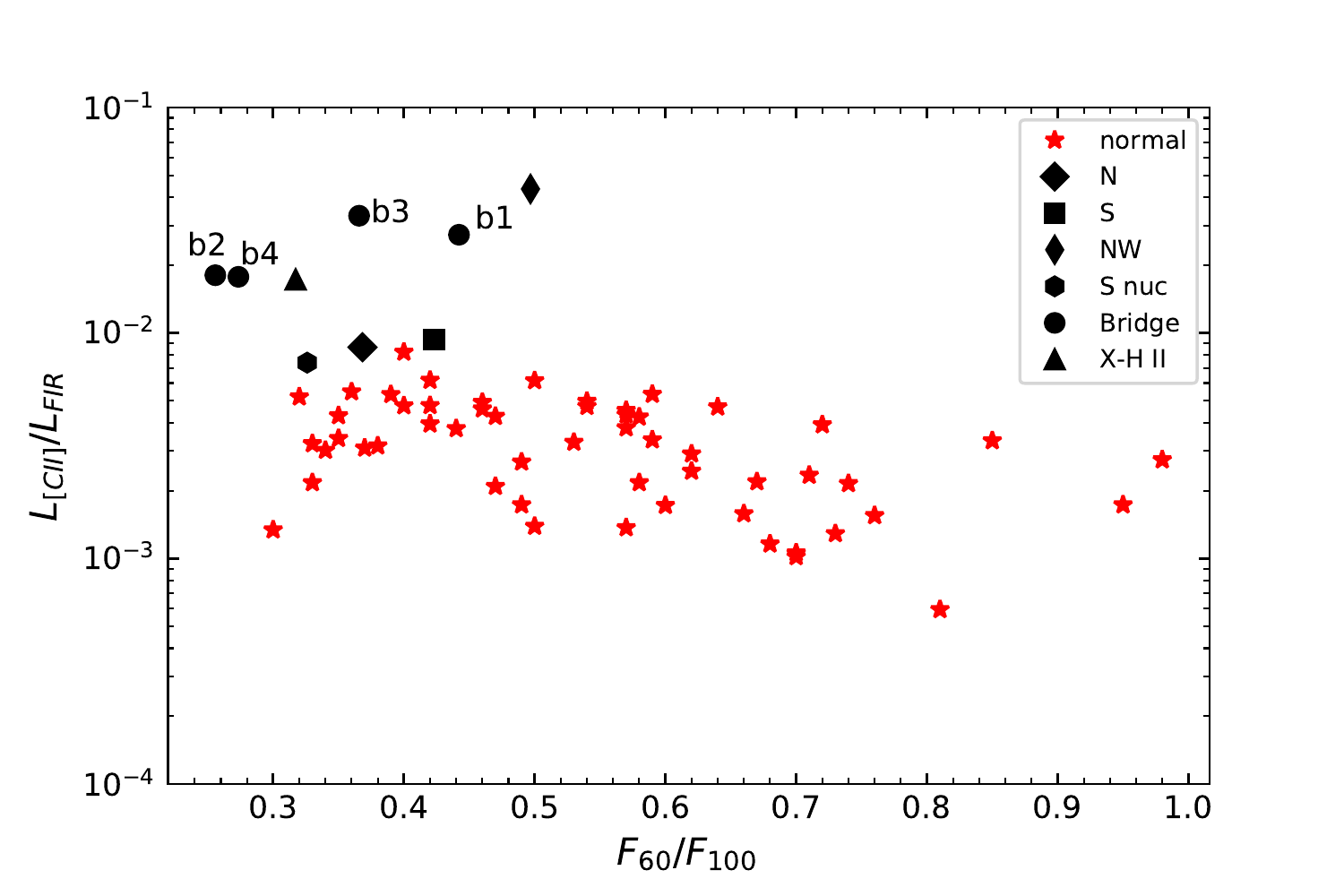}
\caption{[\ion{C}{2}]/FIR ratio vs $F_{\nu}$(60~$\micron$)/$F_{\nu}$(100~$\micron$) (black circles), extragalactic \ion{H}{2} region (black triangle), southern galaxy (black square), southern nucleus (black hexagon), and northern galaxy (black diamond). For comparison, we show normal galaxies from \citet{malhotra2001} (red stars)}. \label{fig:ciiToFIR_color}
\end{figure} 

We can also compare the Taffy results with commonly measured properties of galaxies, such as the far-IR luminosity and color temperature. 
Figure~\ref{fig:ciiToFIR} shows $L_{[C II]}/L_{FIR}$ plotted against $L_{FIR}$ for the Taffy extraction regions. For comparison, we plot normal star-forming galaxies from \citet{malhotra2001} and regions from the group-wide shock in SQ \citep{appleton2013}. We also show nuclei and extended emission from luminous infrared galaxies (LIRGs) in the GOALS sample \citep{diazsantos2013, diazsantos2014}. The full-galaxy extractions and the UGC 12914 nucleus fall at the upper end of $L_{[C II]}/L_{FIR}$ values for star-forming galaxies and LIRGs, so the emission from these regions is likely to be PDR-related. All of the bridge regions, the X-\ion{H}{2} region, and the extended emission from the northwest region of UGC 12915 fall well above the distribution of normal galaxies, again supporting an enhancement in the \ion{C}{2} bridge emission. \\

A similar conclusion is reached by plotting $L_{[C II]}/L_{FIR}$ against the dust temperature as measured by $F_{60 \micron}/F_{100 \micron}$, as shown in Figure~\ref{fig:ciiToFIR_color}. Again, the full-galaxy extractions and UGC 12914 nucleus have $L_{[C II]}/L_{FIR}$ ratios consistent with normal star forming galaxies. The bridge regions, X-\ion{H}{2}, and northwest region of UGC 12915 are all well above the values of normal galaxies in the \citet{malhotra2001} sample. The low $F_{60 \micron}/F_{100 \micron}$ temperatures in the bridge show that the dust is cool there, compared with warm gas temperatures. This is typical of shock-heated regions where the gas is heated more efficiently  that the dust grains -- a result that is contrary to that of a star forming PDR, where much of the heating from young stars goes into warming the grains.     \\

\subsection{[\ion{C}{2}]] boosting from collisional excitation with both \ion{H}{1} and warm H$_2$} \label{sec:hiheat}

Since the Taffy bridge contains large amounts of \ion{H}{1}, it is possible that \ion{H}{1} could contribute to the collisional excitation of [\ion{C}{2}] \citep{Madden1997, Nikola1998}. Indeed, as we have shown in $\S$3.3, \ion{H}{1} alone dominates the low-velocity component of the bridge, with little detected CO emission at those velocities. Although we cannot rule out ``dark'' molecules at those velocities \citep{wolfire2010}, it is worth exploring whether \ion{H}{1} can provide part of the explanation for the [\ion{C}{2}] emission. The surface intensity of [\ion{C}{2}], $I_{C+}$ is given by the equation, \\

\noindent
$I_{C+}$=$h\nu A$/4$\pi$[2e$^{-91/T}$/(1+2e$^{-91/T}$+$n_{Hcr}$/$n_H$)]$\chi$$_{C+}$N$_H$$\Phi$$_b$, \\ 

\noindent
where $h$ is Planck's constant, $A$ is the Einstein coefficient for the transition (we assume $2.29 \times 10^{-6}$~s$^{-1}$), $T$ is the temperature, $n_H$ is the density of \ion{H}{1}. The critical density of \ion{C}{2} for collisions with \ion{H}{1}, $n_{Hcr}$, is a slowly decreasing function of temperature \citep{Goldsmith2012}, and is $\sim$3000~cm$^{-3}$ at $T = 100$~K. $\chi_{C+}$ is the fractional abundance of carbon. If the carbon is evenly split between the neutral and ionized forms, and assuming solar abundance, we adopt $\chi_{C+}$ $\sim$ 1.5$\times$10$^{-4}$. $N_H$ is the column density of neutral hydrogen, and $\Phi_b$ is the beam filling factor which we assume is unity. The assumption of solar abundance in the bridge is based on the belief that most of the gas in the bridge originated in the galaxies before the collision.  

Clearly the density and temperature of the \ion{H}{1} are not constrained by the observations, and so we use the equation to provide an order of magnitude estimate of the importance of collisions with \ion{H}{1}. If we take the peak value of $N_H$ = $1.13 \times$10$^{21}$~cm$^{-2}$ \citep{condon1993}, $T = 100$~K , $n_H$ = 100~cm$^{-2}$ (assuming a cold neutral medium), and with the assumptions given above, we find $I_{C+}$= $0.98 \times 10^{-5}$~erg~s$^{-1}$~cm$^{-2}$~sr$^{-1}$. This is close to the observed peak [\ion{C}{2}] surface density in the bridge at position b3 of $1.3 \times 10^{-5}$~erg~s$^{-1}$~cm$^{-2}$~sr$^{-1}$. Dropping the \ion{H}{1} temperature to 50~K (given that the bridge is likely in adiabatic expansion as the galaxies pull away) would decrease I$_{C+}$ by a factor of 3. Alternatively, increasing the \ion{H}{1} density by a factor of two would more than double the value of $I_{C+}$. Given the large uncertainties in the many assumed quantities, including the filling factor, we can conclude that reasonable properties for the \ion{H}{1} could potentially give rise to a significant fraction of the observed [\ion{C}{2}] emission in the bridge. 

However, we know that \ion{H}{1} is not the only likely collisional partner with ionized carbon atoms. We already know from the previous {\it Spitzer} observations that the warm molecular gas is strongly radiating, so these molecules, which are likely to be shock-heated \citep{peterson2012}, must also play a role in exciting the [\ion{C}{2}] transition\footnote{There could also be an unknown quantity of colder "dark" molecules, which do not radiate strongly in CO, but may be mixed with the \ion{H}{1}. In the present paper we do not include such a component.}. Unlike the case of the \ion{H}{1}, we have better constraints on the properties of the warm H$_2$ emission.  For example, we know the temperature and column density of the bulk of the gas from fitting to the H$_2$ excitation diagram for the pure rotational lines. Adopting a temperature for the warm H$_2$ of 160~K, a density of 1000~cm$^{-3}$ (by analogy with the similar gas in Stephan's Quintet), a molecular column density of $3.2 \times 10^{20}$~cm$^{-2}$, and a critical density of 5900~cm$^{-3}$, the expected intensity of \ion{C}{2} from this collisionally-excited warm H$_2$ would be $3.1 \times$ 10$^{-5}$~erg~s$^{-1}$~cm$^{-2}$~sr$^{-1}$. This is clearly more than enough to explain the observed [\ion{C}{2}] intensity. We conclude that, within the considerable uncertainties, both \ion{H}{1} and warm H$_2$ are likely drivers of the enhanced [\ion{C}{2}] emission seen in the bridge.

\section{Conclusions and Future Directions}
\label{sec:summary}


\vspace{0.3cm}
\begin{itemize}

\item{With the {\it Herschel} PACS and SPIRE spectrometers, we have mapped emission from [\ion{C}{2}]157.7~$\micron$ and [\ion{O}{1}]63.2~$\micron$ in both the galaxies and the bridge. In the bridge we also detected [\ion{C}{1}]2--1 and 1--0 emission, and CO emission lines from the high-J CO transition of $J=$ 6--5, 5--4, and possibly $J=$ 4--3.  The ratio of the [\ion{C}{1}] lines suggests a gas temperature of 35 K for one component of the bridge, if these lines are optically thin. Taken together with the previous detection of a significant mass of warm H$_2$ from pure-rotational mid-IR emission from the bridge with higher temperatures (150 $<$ T $<$ 170 K), and low dust temperatures, the results suggest that a warm, turbulent, multi-phase gaseous medium is present in the bridge which is the result of turbulent motions caused by the recent collision of the two galaxies.} 

\item{The [\ion{C}{2}] and [\ion{O}{1}] emission in the bridge contains multiple kinematically-broad components with intrinsic line widths of 150--250~km~s$^{-1}$ FWHM. The \cii line shape in some parts of the bridge matches both the high- and low-velocity components of the \ion{H}{1}, previously detected by the VLA, whereas the CO (1--0) line is most often associated with the highest-velocity [\ion{C}{2}] component. This implies that the [\ion{C}{2}] emitting gas is distributed within both the \ion{H}{1} and the H$_2$ components in the bridge.}

\item{The [\ion{C}{2}] in the bridge (and part of the northern disk of UGC 12915) appears stronger than expected based on the very low levels of star formation observed there. For example, the [\ion{C}{2}] emission in the bridge deviates from the relationships found between SFR, $L_{IR}$ and [\ion{C}{2}] emission in normal galaxies, suggesting an additional boosting of the emission.} 

\item{We explain the boosted values of [\ion{C}{2}] emission in the bridge as due to a likely combination of both turbulently heated  warm H$_2$ (detected previously by {\it Spitzer}), as well as collisional excitation resulting from the large surface density of neutral hydrogen found in the bridge. With the current data, we cannot determine which of the two processes is dominant in the bridge, although both are likely to be important. Combined, these two collisional partners can excite the [\ion{C}{2}]/FIR ratio to values as high as 3.3$\%$ in regions of the bridge. The failure to detect [\ion{N}{2}]205~$\micron$ in the bridge to a low level, compared with the strength of the [\ion{C}{2}] emission, rules out ionized gas (specifically electrons) as a significant factor in the excitation of the [\ion{C}{2}] transition.}

\end{itemize}

The detection of [\ion{C}{2}], [\ion{O}{1}] and [\ion{C}{1}] emission in the bridge presented here, taken together with the existence of powerful mid-IR warm H$_2$ rotational emission from previous {\it Spitzer} observations, leads to the conclusion that the bridge contains several warm multi-phase gas components, despite having little ongoing star formation. In particular, the cooling time for the rotationally excited H$_2$ is extremely short ($\sim$1000~yrs), and so shock and turbulence must continue to inject energy into the gas to maintain the observed line luminosity \citep[see ][]{peterson2012}. As we have shown, at least part of this energy helps to collisionally excite the [\ion{C}{2}] emission.

One way in which the Taffy system differs from other known examples of turbulent energy dissipation in extragalactic systems, is that the main event that apparently triggered the formation of the bridge was the head-on collision between two gas-rich galaxies some 25~Myr ago. In other examples of similar heating of gas via shocks and turbulence, the source of the turbulence continues to operate -- for instance, the ongoing galaxy collision in Stephan's Quintet, or the propagation of a radio jet into the ISM of a galaxy \citep{appleton2006, cluver2010, guillard2009, appleton2017, ogle2007, ogle2010, guillard2012b}. In the Taffy collision, how can this singular event continue to excite shocks in the bridge so long after the collision?

One solution might be large-scale coherent motions imparted on the gas at the time of the collision.  Models of turbulent energy dissipation \citep[e.g., ][]{Godard2014} suggest that a significant fraction of turbulent energy can be dissipated in large-scale vortices as well as compressive shocks. It is possible that the conditions in the head-on collision between two counter-rotating disks might have led to large-scale coherent vortices on the scale of a few kiloparsecs which may be able to continue to drive shocks throughout the bridge. Although they don't discuss this specifically, the \citet{vollmer2012} models show large-scale structure in the bridge that preserves some of the original rotational motion of the colliding galaxies. Preliminary results from our ALMA mapping of the CO emission in the Taffy bridge support this idea (Appleton et al, in preparation). If such large-scale motions can last long enough, they could provide the source of the heating in the bridge, as density irregularities within large vortices could drive shocks into the surrounding lower-density material.    

Further tests of this idea will require correlating regions of large-scale coherent motions with small scale heating. The combination of millimeter telescope facilities like the NOEMA interferometer and ALMA, for mapping the detailed structure of the denser molecular gas on 100~pc to kiloparsec scales, with observations of warm molecular hydrogen with the {\it James Webb Space Telescope}, will likely provide new insight into how the gas is heated in the Taffy bridge and other similar systems. Furthermore, if small scale shocks provide the main source of dissipation, at the end-point of a turbulent cascade, then the chemistry within the gas may show large scale inhomogenities. For example, as shown by \citet{Godard2014}, chemical reactions between H$_2$ and [\ion{C}{2}] ions can change the relative abundance of [\ion{C}{2}] and CH$^+$. Systems like Taffy and Stephan's Quintet may be remarkable laboratories for future studies of how turbulence on kpc scales can drive non-equilibrium thermal and chemical processes in diffuse intergalactic gas.


\acknowledgments
Acknowledgments. This work is based on observations made with {\it Herschel}, a European Space Agency Cornerstone Mission with significant participation by NASA. Support for this work was provided by NASA through an award issued by JPL/Caltech. BWP also thanks the UW-BC Foundation for its generous support.
UL acknowledges support by the research projects AYA2014-53506-P from the Spanish Ministerio de Econom\'\i a y Competitividad, from the European Regional Development Funds (FEDER) and the Junta de Andaluc\'ia (Spain) grants FQM108. NL acknowledges support by National Key R\&D Program of China, $\#$2017YFA0402704 and NSFC $\#$11673028. YG acknowledges support by National Key R\&D Program of China (2017YFA0402704), the NSFC $\#$11420101002 and the CAS Key Research Program of Frontier Sciences.




\clearpage
\begin{deluxetable}{lrrrr}
\tablecolumns{5}
\tablewidth{0pc}
\tablecaption{PACS Spectral Extraction Regions \label{tab:coords}}
\tablehead{
\colhead{Region} & 
\colhead{R.A.\tablenotemark{a}} &
\colhead{Decl.\tablenotemark{a}} &
\colhead{Size} 
\\
\colhead{} &
\colhead{(J2000.0)} & 
\colhead{(J2000.0)} & 
\colhead{}\\
}
\startdata
UGC 12914 nuc & 0 01 38.\arcsec42 & +23 29 00.\arcsec53 & 18.\arcsec3 & 16.\arcsec3\\
UGC 12915 NW\tablenotemark{b} & 0 01 39.\arcsec43 & +23 30 09.\arcsec28 & 19.\arcsec1 & 20.\arcsec0\\
b1 & 0 01 38.\arcsec42 & +23 29 48.\arcsec73 & 28.\arcsec0 & 16.\arcsec8\\
b2 & 0 01 39.\arcsec33 & +23 29 31.\arcsec83 & 19.\arcsec4 & 17.\arcsec0\\
b3 & 0 01 40.\arcsec70 & +23 29 32.\arcsec16 & 18.\arcsec3 & 16.\arcsec3\\
b4 & 0 01 41.\arcsec47 & +23 28 58.\arcsec34 & 22.\arcsec4 & 16.\arcsec8\\
X-\ion{H}{2} & 0 01 40.\arcsec66 & +23 29 15.\arcsec14 & 33.\arcsec6 & 16.\arcsec8\\
\enddata
\tablenotetext{a}{Coordinates of center of extraction box.}
\tablenotetext{b}{NW region was also rotated by 46.6$\degr$.}
\end{deluxetable}

\begin{deluxetable*}{lrrrrrr}
\tablecolumns{7}
\tablewidth{0pc}
\tablecaption{SED Fit Properties \label{tab:sed}}
\tablehead{
\colhead{Region} & 
\colhead{$M_{stars}$} &
\colhead{$M_{dust}$} &
\colhead{SFR} &
\colhead{log($L_{TIR}/L_{\sun}$) \tablenotemark{a}} &
\colhead{log($L_{FIR}/L_{\sun}$) \tablenotemark{b}} &
\colhead{$F_{60}/F_{100}$\tablenotemark{c}}
\\
\colhead{} &
\colhead{$10^9 M_{\sun} $} & 
\colhead{$10^6 M_{\sun} $} &
\colhead{$M_{\sun}$ yr$^{-1}$} & 
\colhead{} &
\colhead{} &
\colhead{}  \\
}
\startdata
UGC 12915 & 43.65 $\pm$ 4.21 & $40.46\pm 6.41$ & $2.77 \pm 0.30$ & $10.74 \pm 0.06$ & 10.41 $\pm$ 0.06 & 0.37 \\
UGC 12914 & 77.62 $\pm$ 11.51 & $40.93 \pm 6.49$ & 1.13 $\pm$ 0.01 & 10.49 $\pm$ 0.03 & 10.09 $\pm$ 0.03 & 0.42 \\
South Nucleus & 26.3 $\pm$ 9.18 & $5.15 \pm 0.23$ & 0.07 $\pm$ 0.01 & 9.84 $\pm$ 0.04 & 9.45 $\pm$ 0.04 & 0.33 \\
UGC 12915 NW & 2.49 $\pm$ 0.27 & $2.28 \pm 0.59$ & 0.03 $\pm$ 0.01 & 8.99 $\pm$ 0.03 & 8.51 $\pm$ 0.03 & 0.50 \\
b1 & 1.42 $\pm$ 0.02 & $0.35 \pm 0.05$ & 0.04 $\pm$ 0.01 & 8.88 $\pm$ 0.01 & 8.53 $\pm$ 0.01 & 0.44 \\
b2 & 2.00 $\pm$ 0.32 & $4.75 \pm 0.84$ & 0.04 $\pm$ 0.01 & 9.15 $\pm$ 0.02 & 8.66 $\pm$ 0.02 & 0.26 \\
b3 & 2.41 $\pm$ 0.26 & $2.44 \pm 0.69$ & 0.04 $\pm$ 0.01 & 8.99 $\pm$ 0.02 & 8.52 $\pm$ 0.02 & 0.37 \\
b4 & 1.34 $\pm$ 0.03 & $1.00 \pm 0.41$ & 0.01 $\pm$ 0.01 & 8.59 $\pm$ 0.01 & 8.08 $\pm$ 0.01 & 0.27 \\
X-\ion{H}{2} & 1.96 $\pm$ 0.14 & $4.98 \pm 1.11$ & 0.09 $\pm$ 0.01 & 9.40 $\pm$ 0.01 & 9.01 $\pm$ 0.01 & 0.32 \\
\enddata
\tablenotetext{a}{3--1000~\micron}
\tablenotetext{b}{42--122~\micron}
\tablenotetext{c}{$F_{\nu}$(60~$\micron$)/$F_{\nu}$(100~$\micron$), where $F_{\nu}$ is in Jy}
\end{deluxetable*}

\begin{deluxetable}{lccccccccc}
\tablecolumns{10}
\tablewidth{0pc}
\tablecaption{PACS \cii Measurements \label{tab:cii}}
\tablehead{
\colhead{Region} & 
\colhead{Line Flux} &
\colhead{$V_1$} &
\colhead{$\Delta$V$_{1}$\tablenotemark{a}}  &
\colhead{$\Delta$V1$$\tablenotemark{b}} &
\colhead{$V_1$ frac\tablenotemark{c}} &
\colhead{$V_2$} &
\colhead{$\Delta$V$_{2}$\tablenotemark{a}} &
\colhead{$\Delta$V$_{1}$\tablenotemark{b}} &
\colhead{$V_2$ frac\tablenotemark{c}}
\\
\colhead{} &
\colhead{} & 
\colhead{Heliocentric} & 
\colhead{FWHM} &
\colhead{FWHM} &
\colhead{} &
\colhead{Heliocentric} &
\colhead{FWHM} &
\colhead{FWHM} &
\colhead{}
\\
\colhead{} &
\colhead{($10^{-17}$ W m$^{-2}$)} & 
\colhead{(km s$^{-1}$)} & 
\colhead{(km s$^{-1}$)} &
\colhead{(km s$^{-1}$)} &
\colhead{(percent)} &
\colhead{(km s$^{-1}$)} &
\colhead{(km s$^{-1}$)} &
\colhead{(km s$^{-1}$)} &
\colhead{(percent)}
\\
}
\startdata
UGC 12915 & 186.3 $\pm$ 19.1 & 4369 & 369 & 284 & 50 & 4692 & 306 & 195 & 50 \\
UGC 12914 & 96.4 $\pm$ 10.2 & 4188 & 297 & 181 & 52 & 4516 & 287 & 165 & 48 \\
UGC 12914 nuc & 17.3 $\pm$ 1.8 & 4367 & 425 & 354 & 100 & \nodata & \nodata & \nodata & \nodata \\
UGC 12915 NW & 11.6 $\pm$ 1.2 & 4497 & 484 & 423 & 49 & 4676 & 278 & 148 & 51 \\
b1 & 7.7 $\pm$ 1.0 & 4141 & 249 & 82 & 41 & 4509 & 445 & 377 & 59 \\
b2 & 6.9 $\pm$ 0.9 & 4214 & 278 & 149 & 68 & 4508 & 276 & 145 & 32 \\
b3 & 9.1 $\pm$ 1.0 & 4323 & 334 & 237 & 65 & 4533 & 271 & 135 & 35 \\
b4 & 1.8 $\pm$ 0.2 & 4343 & 318 & 215 & 100 & \nodata & \nodata & \nodata & \nodata \\
\ion{H}{2} & 14.8 $\pm$ 1.5 & 4382 & 457 & 392 & 100 & \nodata & \nodata & \nodata & \nodata \\
\enddata
\tablenotetext{a}{FWHM from observed line profile}
\tablenotetext{b}{Intrinsic FWHM after deconvolution with instrument profile}
\tablenotetext{c}{Fraction of total flux in each component.}
\end{deluxetable}

\begin{deluxetable}{lccccccccc}
\tablecolumns{10}
\tablewidth{0pc}
\tablecaption{PACS \oi Measurements \label{tab:oi}}
\tablehead{
\colhead{Region} & 
\colhead{Line Flux} &
\colhead{$V_1$} &
\colhead{$\Delta$V$_{1}$\tablenotemark{a}}  &
\colhead{$\Delta$V1$$\tablenotemark{b}} &
\colhead{$V_1$ frac\tablenotemark{c}} &
\colhead{$V_2$} &
\colhead{$\Delta$V$_{2}$\tablenotemark{a}} &
\colhead{$\Delta$V$_{1}$\tablenotemark{b}} &
\colhead{$V_2$ frac\tablenotemark{c}}
\\
\colhead{} &
\colhead{} & 
\colhead{Heliocentric} & 
\colhead{FWHM} &
\colhead{FWHM} &
\colhead{} &
\colhead{Heliocentric} &
\colhead{FWHM} &
\colhead{FWHM} &
\colhead{}
\\
\colhead{} &
\colhead{($10^{-17}$ W m$^{-2}$)} & 
\colhead{(km s$^{-1}$)} & 
\colhead{(km s$^{-1}$)} &
\colhead{(km s$^{-1}$)} &
\colhead{(percent)} &
\colhead{(km s$^{-1}$)} &
\colhead{(km s$^{-1}$)} &
\colhead{(km s$^{-1}$)} &
\colhead{(percent)}
}
\startdata
UGC 12915 & 67.7 $\pm$ 8.4 & 4288 & 169 & 146 & 27 & 4516 & 165 & 142 & 73 \\
UGC 12914 & 37.4 $\pm$ 4.1 & 4188 & 297 & 181 & 56 & 4542 & 149 & 123 & 44 \\
UGC 12914 nuc & 7.8 $\pm$ 0.8 & 4370 & 300 & 288 & 100 & \nodata & \nodata & \nodata & \nodata \\
UGC 12915 NW & 4.3 $\pm$ 0.5 & 4598 & 215 & 198 & 100 & \nodata & \nodata & \nodata & \nodata \\
b1 & $\leqslant1.47$ & \nodata & \nodata & \nodata & \nodata & \nodata & \nodata & \nodata & \nodata \\
b2 & 5.0 $\pm$ 0.6 & 4312 & 286 & 273 & 100 & \nodata & \nodata & \nodata & \nodata \\
b3 & 3.3 $\pm$ 0.7 & 4234 & 255 & 99 & 100 & \nodata & \nodata & \nodata & \nodata \\
b4 & $\leqslant3.0$  & \nodata & \nodata & \nodata & \nodata & \nodata & \nodata & \nodata & \nodata \\
\ion{H}{2} & 7.8 $\pm$ 0.8 & 4267 & 176 & 155 & 25 & 4518 & 165 & 142 & 75 \\
\enddata
\tablenotetext{a}{FWHM from observed line profile}
\tablenotetext{b}{Intrinsic FWHM after deconvolution with instrument profile}
\tablenotetext{c}{Fraction of total flux in each component.}
\end{deluxetable}

\begin{deluxetable}{lccc}
\tablecolumns{4}
\tablewidth{0pc}
\tablecaption{Line Ratios \label{tab:linerats}}
\tablehead{
\colhead{Region} & 
\colhead{[\ion{C}{2}]/[\ion{O}{1}]} &
\colhead{[\ion{C}{2}]/CO\tablenotemark{a}} &
\colhead{[\ion{C}{2}]/$L_{FIR}$\tablenotemark{b}} 
\\
\colhead{} &
\colhead{} & 
\colhead{} & 
\colhead{percent} 
\\
}
\startdata
UGC 12915 & $2.75 \pm 0.44$ & $669 \pm 67$ & $0.86 \pm 0.09$ \\
UGC 12914 & $2.58 \pm 0.39$ & $465 \pm 47$ & $0.94 \pm 0.10$ \\
UGC 12914 nuc & $2.23 \pm 0.32$ & $403 \pm 40$ & $0.74 \pm 0.07$ \\
UGC 12915 NW & $2.71 \pm 0.40$ & $1286 \pm 129$ & $4.35 \pm 0.44$ \\
b1 & $\geqslant5.22$ & $909 \pm 91$ &$ 2.73 \pm 0.36$ \\
b2 & $1.39 \pm 0.24$ & $348 \pm 35$ & $1.80 \pm 0.24$ \\
b3 & $2.75 \pm 0.66$ & $218 \pm 22$ & $3.31 \pm 0.35$ \\
b4 & $\geqslant0.59$ & $191 \pm19 $ & $1.78 \pm 0.20$ \\
X-\ion{H}{2} & $1.90 \pm 0.27$ & $236 \pm 24$ & $1.72 \pm 0.18$ \\
\enddata
\tablenotetext{a}{The PACS \ion{C}{2} and BIMA CO beams have comparable sizes of $9.\arcsec4 \times 9.\arcsec4$ and $9.\arcsec9 \times 9.\arcsec7$, respectively. The uncertainties in the CO fluxes were assumed to be 10\%.}
\tablenotetext{b}{42--122 \micron}
\end{deluxetable}

\begin{deluxetable}{lccccc}
\tablecolumns{6}
\tablewidth{0pc}
\tablecaption{SPIRE FTS Line Fluxes from Taffy Bridge\tablenotemark{a}\label{tab:spire}}
\tablehead{
\colhead{Line} & 
\colhead{$\nu_{obs}$} &
\colhead{FWHM\tablenotemark{b}} &
\colhead{V$_{helio}$ (unc)} &
\colhead{Line Flux (unc)} &
\colhead{Detector}  
\\
\colhead{} & 
\colhead{(GHz)} &
\colhead{(arcsec)} &
\colhead{(km s$^{-1}$) } &
\colhead{(x 10$^{-17}$W m$^2$)} &
\colhead{} 
}
\startdata
 12CO 4-3 & 453.93 & 43 & $4699 \pm 28$ & [$1.33 \pm 0.09$]\tablenotemark{c} & SLWC3 \\
 12CO 5-4 & 567.80 & 35 & $4473 \pm 44$ & $0.68 \pm 0.08$ & SLWC3 \\
 12CO 6-5 & 681.09 & 32& $4571 \pm 76$ & $0.33 \pm 0.09$ & SLWC3 \\
 $[$\ion{C}{1}$]$ 1-0 & 484.74 & 38 & $4591 \pm 29$ & $1.20 \pm 0.09$  & SLWC3 \\
$[$\ion{C}{1}$]$ 2-1 & 797.50 & 35 & $4452 \pm 29$ & $0.80 \pm 0.09$ & SLWC3 \\
$[$\ion{N}{2}$]$205~$\micron$ & 1440.78 & 18 & \nodata  & $<$ 0.08\tablenotemark{d} & SSWD4 \\
\enddata
\tablenotetext{a}{Observations centered on $\alpha$(J2000) = 00hr 01m 40.3s, $\delta$(J2000) = 23d 29m 22.0s.}
\tablenotetext{b}{The dimensions of the SLW FWHM FTS beam is not proportional to $\nu$$^{-1}$ and is somewhat non-circular because of the design of the feed-horn (see Figure 5.18 of the SPIRE Observer's Manual;    \\        
$http://herschel.esac.esa.int/Docs/SPIRE/html/spire\_om.html$)}
\tablenotetext{c}{Uncertain detection because the line is close to the SLW lower band-edge.}
\tablenotetext{d}{3$\sigma$ upper limit.}

\end{deluxetable}


\end{document}